\documentclass[12pt,preprint]{aastex}

\slugcomment{ }

\shorttitle{Collapsed Cores in Globular Clusters}
\shortauthors{Watanabe et al.}

\begin{document}


\title{An Unbiased Spectral Line Survey toward R~CrA~IRS7B in the 345~GHz Window with ASTE}


\author{Yoshimasa~Watanabe\altaffilmark{1}, Nami~Sakai\altaffilmark{1}, Johan~E.~Lindberg\altaffilmark{2,3}, Jes~K.~J{\o}rgensen\altaffilmark{3,2}, Suzanne~E.~Bisschop\altaffilmark{2,3} and Satoshi~Yamamoto\altaffilmark{1}}


\altaffiltext{1}{Department of Physics, The University of Tokyo, 7-3-1 Hongo, Bunkyo-ku, Tokyo, 113-0033, Japan}
\altaffiltext{2}{Centre for Star and Planet Formation, Natural History Museum of Denmark, University of Copenhagen, {\O}ster Voldgade 5-7, DK-1350 Copenhagen K., Denmark}
\altaffiltext{3}{Niels Bohr Institute, University of Copenhagen, Juliane Maries Vej 30, DK-2100 Copenhagen {\O}., Denmark}

\begin{abstract}
 We have conducted a spectral line survey in the 332 - 364~GHz region with the ASTE~10~m telescope toward R~CrA~IRS7B, a low-mass protostar in the Class 0 or Class 0/I transitional stage.  We have also performed some supplementary observations in the 450~GHz band.  In total, 16 molecular species are identified in the 332 - 364~GHz region. Strong emission lines of CN and CCH are observed, whereas complex organic molecules and long carbon-chain molecules which are characteristics of hot corino and warm carbon-chain chemistry (WCCC) source, respectively, are not detected.  The rotation temperature of CH$_{3}$OH is evaluated to be 31~K, which is significantly lower than that reported for the prototypical hot corino IRAS~16293-2422 ($\sim$85~K).  The deuterium fractionation ratios for CCH and H$_{2}$CO are obtained to be 0.038 and 0.050, respectively, which are much lower than those in the hot corino.  These results suggest a weak hot corino activity in R~CrA~IRS7B.  On the other hand, the carbon-chain related molecules, CCH and c-C$_3$H$_2$, are found to be abundant.  However, this source cannot be classified as a WCCC source, since long carbon-chain molecules are not detected.  If WCCC and hot corino chemistry represent the two extremes in chemical compositions of low-mass Class 0 sources, R~CrA~IRS7B would be a source with a mixture of these two chemical characteristics.  The UV radiation from the nearby Herbig Ae star R~CrA may also affect the chemical composition.  The present line survey demonstrates further chemical diversity in low-mass star-forming regions.  
\end{abstract}


\section{Introduction}
One of the key issues in low-mass star formation studies is to observationally address how the interstellar matter is brought into a protoplanetary disk.  This is of fundamental importance, since it is eventually related to the origin of rich materials in our solar system.  A great advance toward this direction in the last decade was the recognition of warm inner regions around low-mass protostars \citep{bla94,dis95,cec98,cec20,sch02}, and the detection of various complex organic molecules including HCOOCH$_3$, (CH$_3$)$_2$O, and C$_2$H$_5$CN in four Class 0 protostars, IRAS~16293-2422, NGC~1333~IRAS4A, IRAS4B, and IRAS2A \citep{caz03,bot04,bot07,kua04,jor05a,cha05,sak06,bis08}.  Although these complex organic molecules had long been recognized as molecules which are characteristic to hot cores in high-mass star-forming regions, it is now evident that some low-mass star-forming regions harbor them in a high temperature ($>$ 100~K) and high density ($>$ $10^6$~cm$^{-3}$) region around a protostar, called hot corino.  So far, only four hot corinos are known and three of them belong to the Perseus region, and hence, whether the hot corino phenomenon is common among protostars is still unclear  (See the review by \citet{herb09}).

As a counter example, \citet{sak08,sak09a} found that some Class 0 objects have completely different chemical compositions.  Prototypical examples are L1527 in Taurus and IRAS~15398-3359 in Lupus, where various carbon-chain molecules such as C$_4$H, C$_4$H$_2$, and HC$_5$N are extremely abundant.  In contrast to the hot corino case, the complex organic molecules are deficient in these sources.   Carbon-chain molecules are thought to be produced from CH$_4$ evaporated from grain mantles by protostellar activities (Warm Carbon-Chain Chemistry;~WCCC).  The known sources showing hot corino and WCCC characteristics are, however, all Class 0 sources according to their bolometric temperature \citep{eva09}.  The two groups of sources are therefore also not necessarily just protostars in different evolutionary stages.

This discussion also illustrates the chemical diversity of protostars: some associated with hot corinos, some showing molecules related to the WCCC activities, and some neither.  One interpretation of such a chemical diversity is that the hot corinos and the WCCC sources may represent the two extremes with respect to chemical compositions of Class 0 young stellar objects.  If so, there should exist a protostellar core that possesses the intermediate characteristics.  \citet{sak09a} carried out a C$_4$H survey toward 17 low-mass protostars, and found that the C$_4$H abundance varies from source to source, where it is highest for the WCCC sources and lower for hot corinos.  Sources between the two extremes indeed exist with respect to the C$_4$H abundance.  They proposed that the variation would originate from difference of the duration time of the starless-core phase.  Alternatively, the other environmental effects such as UV radiation from nearby stars would also contribute to the chemical variation.  Understanding an origin of the chemical diversity remains an important task.  A particularly important method for this is an unbiased spectral line survey.

Spectral line surveys have been carried out toward representative sources with strong emission lines, such as high-mass star-forming regions \citep[e.g.~Orion KL;][]{bla86,sch97}, envelopes of late-type stars \citep[e.g.~IRC+10216;][]{ave92}, and active nuclei in galaxies \citep[e.g.~NGC 253;][]{mat06}.  On the other hand, spectral line surveys toward low-mass star-forming sources and starless cores are relatively sparse.  This is because unbiased line surveys for such regions need a lot of observation time due to weak and narrow emission lines.  The exceptions are line surveys of the hot corino IRAS~16293-2422 \citep{bla94,dis95,cau11} and of the starless core TMC-1 \citep{kai04}.  Except for these two, the chemical compositions of low-mass star-forming regions have so far been studied by targeting only specific molecules \citep[e.g.][]{mar04,jor05b,bot07}.  Recent developments of the ALMA-type low noise receivers and wide-band auto-correlation spectrometers have made wide-band survey observations possible even toward the low-mass star-forming regions.

With this in mind, we have conducted a spectral line survey toward the low-mass protostar R~CrA~IRS7B in the Corona Australis dark cloud at a distance of about 170 pc \citep{kh98}.  The cloud is extended from south to north, lying almost perpendicularly to the galactic plane.  The total mass is estimated to be about 7000 $M_{\odot}$ \citep{cp91}.  Around the Herbig Ae star R~CrA, \citet{haj93} found a dense molecular cloud core (A2) based on the C$^{18}$O observations using the SEST telescope.  In the R~CrA region, IRS7 is the most reddened source \citep{ts84}.  Two continuum peaks, IRS7A and IRS7B, separated by 12$''$ were detected by \citet{bro87} with VLA.  \citet{nut05} identified a Class 0 protostar in IRS7B by the 450 $\mu$m and 850 $\mu$m SCUBA observations, although \citet{gro07} later suggested it to be objet in transition between Class 0 and Class I.  Toward this position, strong emission lines of H$_2$CO and CH$_3$OH were detected \citep{sch06}.  From these results, R~CrA~IRS7B has been thought as a hot corino candidate (Lindberg et~al. in prep.).  In this paper, we will present a spectral line survey toward R~CrA~IRS7B in the 345~GHz band.  A related spectral line survey covering frequencies between 218 and 245~GHz is being conducted with APEX, and will be reported separately (Lindberg et~al. in prep.).

\section{Observations}
Observations toward R~CrA~IRS7B were carried out in June and August 2010 with the ASTE 10~m telescope \citep{eza04}.  The observed position was the SMA 220~GHz continuum peak ($\alpha_{\rm J2000}$, $\delta_{\rm J2000}$) = (19$^{\rm h}$ 01$^{\rm m}$ 56$^{\rm s}$.4 , -36$^{\circ}$ 57$'$ 28$''$.3) (Lindberg et~al. in prep.).  In the 345~GHz band, the beam size is $\sim$22$''$, and the main beam efficiency ($\eta_{\rm mb}$) is $\sim$60~\%.  A side-band separating (2SB) mixer receiver (CATS345) was used as a frontend, whose typical system temperature ranged from 200 to 500~K, depending on the atmospheric conditions.  The backend was a bank of XF-type digital spectro-correlators (MAC), whose bandwidths are 512 MHz, all having 1024 spectral channels.  The frequency resolution is 0.5~MHz, which corresponds to $\sim$0.5~km~s$^{-1}$ at 345~GHz.  This resolution is sufficient, since a typical line width for R~CrA~IRS7B is $\sim$2~km~s$^{-1}$.  A position-switching method was employed with the off-position at ($\Delta\alpha$, $\Delta\delta$) = (+30$'$, 0.0$'$).  The telescope pointing was checked once every hour by observing a bright point-like $^{12}$CO($J=3-2$) source, V5104Sgr.  The pointing accuracy was ensured to be better than 5$''$.  The intensity calibration was carried out by the chopper-wheel method.  The antenna temperature ($T_{\rm A}^{*}$) is converted to the main-beam brightness temperature ($T_{\rm mb}$) by $T_{\rm mb}=T_{\rm A}^{*}/\eta_{\rm mb}$ ($\eta_{\rm mb}=0.6$).  The intensity fluctuation from day to day was about 19~\%, as evaluated from the $^{12}$CO($J$=3-2) intensity of V5104Sgr used for the pointing.  The total observation time was 42 hours to cover the frequency range from 332 GHz to 364 GHz.

In addition to the observations in the 345~GHz band, some supplementary observations were carried out in the 450~GHz band with ASTE in November 2010.  The ALMA Band 8 qualified model receiver was used as a frontend.  Since the observation time was limited (6 hours), a few small frequency ranges covering important lines were observed.  The covered frequency ranges are 435 - 437~GHz, 459.25 - 461.25~GHz, 490.0 - 490.5~GHz, and 491.0 - 492.5~GHz.  The beam size is $\sim$ 17$''$, and the main beam efficiency is $\sim$ 50~\% at 450~GHz.  The same backend as for the 350~GHz observation (MAC) was used, whose velocity resolution at 450~GHz is $\sim$0.3~km~s$^{-1}$.  The $T_{\rm A}^{*}$ scale is converted to the $T_{\rm mb}$ scale by using the main beam efficiency ($\eta_{\rm mb}=0.5$).  A position-switching method was employed with the off position at ($\Delta\alpha$, $\Delta\delta$) = (+30$'$, 0.0$'$), as in the case of the 345~GHz observation.  The intensity calibration was carried out by the chopper-wheel method.  A typical system temperature was 800 - 2000~K, depending critically on the atmospheric condition and observing frequencies.

The observed data were reduced with NEWSTAR, which is a software package developed by NRO.  Spectral baselines were subtracted by fitting a line-free part to the 5th - 7th order polynomial in a frequency range of 500 MHz.  Distorted sub-scan spectra due to bad atmospheric conditions and instabilities of the receiver system, whose baseline could not be subtracted by fitting the polynomial curves, were excluded in the integration procedure.

\section{Results}
\subsection{Overall Feature} \label{overallf}
\begin{figure}
\epsscale{0.95}
\plotone{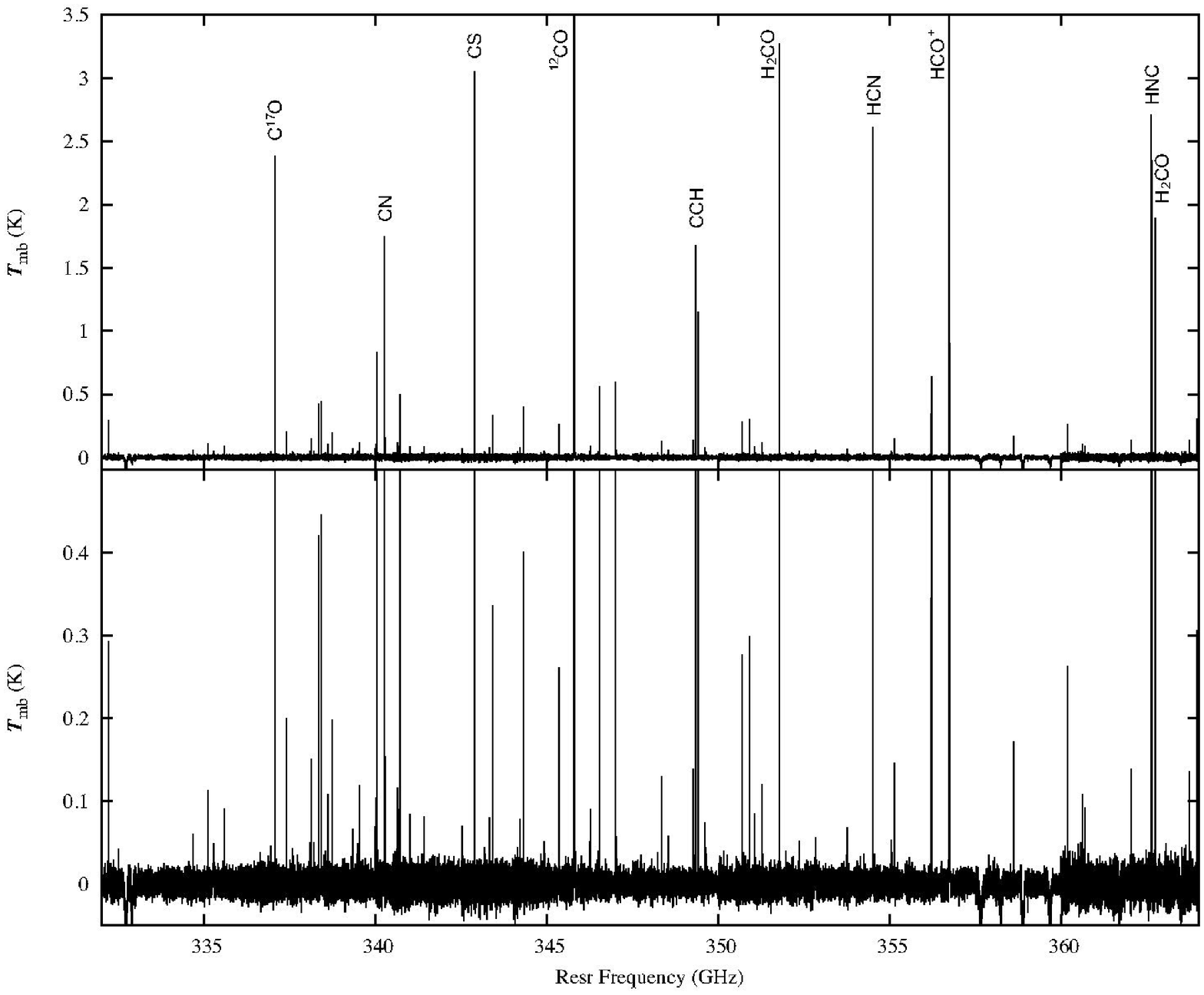}
\caption{Compressed spectrum of R~CrA~IRS7B observed in the 345~GHz band (upper), and its expansion in the vertical scale to show faint lines (lower).  \label{fig1}}
\end{figure}

Figure \ref{fig1} shows the compressed spectrum from 332 to 364~GHz, whereas Figure \ref{fig2} shows its expansions of every 1~GHz interval.  Typically, the r.m.s. noise ranges from 12 to 23 mK in $T_{\rm mb}$.  In some parts of the spectrum, the periodic baseline distortion remains, which actually limits the sensitivity.  'Absorption-like' features with negative intensities can be seen in the spectrum, which are caused by telluric ozone (O$_3$).  In total, 89 emission lines are detected in the frequency range from 332 to 364~GHz (Figure \ref{fig3}; Table \ref{tab1}).  Hence, the line density is 2.8~GHz$^{-1}$ with this sensitivity.  We identified 16 fundamental molecular species and 16 isotopomers with the aid of spectral line databases CDMS \citep{mul01,mul05} and JPL \citep{pic98}.  The $V_{\rm LSR}$ value is assumed to be 6~km~s$^{-1}$.  When identifying weak emission lines, we carefully confirmed them by checking the presence of other lines of the same species at other frequencies.  In the 450~GHz band, 6 emission lines were detected, from which 3 molecular and one atomic species (C) were identified.  The line identification list is given in Table \ref{tab1}, and the individual line profiles are shown in Figure \ref{fig3}. 

All the identified molecules consist of 3 heavy atoms or less.  Sulfur dioxide (SO$_2$) is the heaviest molecule detected in this observation.  Neither complex organic molecules nor long carbon-chain molecules were detected.  Although this seems to be due to low abundances of heavy molecules, another possible reason would be insufficient excitation.  In general, heavy molecules have small rotational constants, and hence, their rotational transitions in the submillimeter-wave region usually have high upper-state energies.  For instance, the $31_{0\,31}-30_{0\,30}$ line of HCOOCH$_{3}$ at 333~GHz has the upper state energy of 259~K.  Similarly, the $J=38-37$ line of HC$_3$N at 346~GHz has the upper state energy of 323~K.  Hence, excitation of such transitions requires high density and high temperature conditions.  If the emitting region is small and the molecular abundance is not very high, the emission of the heavy molecules will hardly be detected in the submillimeter-wave band.  Thus, the excitation condition would seriously limit the detectability of molecular lines, though this can sometimes be a merit of the submillimeter-wave observations in reducing the weed spectral lines to avoid the line confusion problem \citep[e.g.][]{ter10}.  

Very strong emission lines of fundamental molecules such as CO, HCN, HCO$^+$, and H$_2$CO are readily seen in Figure \ref{fig1}.  In addition to them, we found very strong emission lines of CN ($N=3-2$) and CCH ($N=4-3$) in R~CrA~IRS7B.  Ten hyperfine components emission lines were detected for CN with the peak intensity of 2.9~K in $T_{\rm mb}$, whereas 7 hyperfine component lines were found for CCH with the peak intensity of 2.8~K (Figure \ref{fig3}).  Furthermore, 8 lines of c-C$_3$H$_2$ were also detected with moderate intensities.  The strongest line of c-C$_3$H$_2$ is $5_{5\,0}-4_{4\,1}$ at 349.264~GHz, whose intensity is 0.23~K (Figure \ref{fig3}).  

The NO $^{2}\Pi_{\frac{1}{2}}$ $J=7/2-5/2$ lines were detected in this survey.  One $\Lambda$-type doubling component is clearly seen, which is split into three hyperfine component lines.  Another $\Lambda$-type doubling component is partly blended with the CH$_3$OH line.  The NO molecule has been found in Sgr B2 \citep{lis78}, the high-mass star-forming region OMC-1 \citep{woo84}, and cold dark clouds like L134N and TMC-1 \citep{mcg90}.  This molecule is a reaction intermediate to form N$_2$, and is one of key molecules to understand the nitrogen chemistry in the gas phase.

As for the sulfur bearing species, three bright lines of SO, one line of CS, 2 lines of H$_2$CS, one line of HCS$^{+}$, and 9 lines of SO$_2$ were detected.  Furthermore, the SO$^+$ $J=15/2 - 13/2$ lines were possibly detected.  Two $\Lambda$-type doubling components of SO$^{+}$ are recognizable at 347.470~GHz and 348.115~GHz with the confidence levels of 4$\sigma$ and 3$\sigma$, respectively, although the lines suffer from the baseline distortion because of their low intensities, particularly for the lower $\Lambda$-type doubling component (Figures \ref{fig2} and \ref{fig3}).  So far SO$^+$ has been detected in the shocked cloud associated with the supernova remnants \citep[IC443G;][]{tur92} and photodissociation regions (PDRs) \citep[Orion Bar;][]{fue03} as well as some high-mass and low-mass star-forming regions \citep{woo87,tur94,sta07}.

In addition, we notice a few important non-detections.  The $J=3-2$ lines of CO$^+$, which is thought to be abundant in PDRs, fall in the observed frequency range, but were not detected.  Furthermore, the HOC$^+(J=4-3)$ line was not found in this survey.  The linear isomer of c-C$_3$H$_2$, l-C$_3$H$_2$, was not detected.  Although (CH$_3$)$_2$O has a relatively low excitation line ($5_{5\,0}-4_{4\,1}$, E$_{\rm u}$ = 48.8~K) at 358.452~GHz, it was not visible in the present observations.  

\subsection{Rotation Temperatures and Column Densities} \label{TempDens}
Assuming optically thin emission and local thermodynamic equilibrium (LTE), we determined the rotation temperatures and the beam-averaged column densities of CH$_3$OH, SO$_2$, and c-C$_3$H$_2$ from the 345~GHz band data by using the rotation diagram method (Table \ref{tab2} and Figure \ref{fig4}).   As for c-C$_3$H$_2$, a special treatment is needed.  Two of the 8 observed c-C$_{3}$H$_{2}$ lines (351.782~GHz and 351.66~GHz) are composites of the ortho and para c-C$_3$H$_2$ lines with almost the same upper state energies.  We included these two composite lines in the rotation diagram analysis by assuming the ortho-to-para ratio of 3.  Here, we also assumed the same rotation temperature for the ortho and para c-C$_3$H$_2$.  As for CCH and H$_{2}$CO, we evaluated their rotation temperatures from the 345~GHz and 450~GHz band data, as shown in Table \ref{tab2} and Figure \ref{fig4}, where the beam filling factor is assumed to be the same for the both bands.  

As summarized in Table \ref{tab2}, the rotation temperatures of c-C$_3$H$_2$ CH$_3$OH, SO$_2$, CCH, and H$_2$CO range from 16~K to 31~K, being almost comparable to one another.  The rotation temperatures are significantly higher than those ($<$10~K) found in cold dark clouds \citep[e.g.][]{sak08}.  On the other hand, they are much lower than that found in the typical hot corino IRAS~16293-2422 \citep{dis95}, and comparable to those found for H$_2$CO in low-mass protostellar sources without hot corino activities such as L1157-mm and L1527 \citep[18~K and 16~K, respectively:][]{mar04}.  It should be noted that the rotation temperatures of c-C$_3$H$_2$ and CCH are similar to those of CH$_3$OH and SO$_2$ in R~CrA~IRS7B, suggesting that all these molecules are subject to the same physical conditions, and thus likely reside in the same region.  Since the rotation temperatures are much lower than the upper state energies of the heavy molecules such as HCOOCH$_{3}$ $(31_{0\,31}-30_{0\,30};~259~{\rm K})$ and HC$_{3}$N $(J=38-37;~323~{\rm K})$, these lines are hardly detected in R~CrA~IRS7B.   

The beam-averaged column densities of the other molecules were evaluated by assuming optically thin emission and LTE, where the excitation temperature was assumed to be 20~K.  In order to see how the derived column densities depend on the assumed excitation temperature, the column densities were also calculated for the excitation temperatures of 15~K and 25~K.  Although the ratio of the column density estimated from the normal species line to that from the $^{13}$C species line is close to the isotope abundance ratio of 60 \citep{luc98} for H$_2$CO, the corresponding ratios for HCO$^{+}$, HCN, and HNC are found to be lower than 60.  This indicates that the bright lines, such as the HCO$^{+}$, HCN, and HNC lines, are not always optically thin.  Therefore, we used the data of the $^{13}$C species to derive the column densities of HCO$^{+}$, HCN, HNC, and H$_2$CO, where the $^{12}$C/$^{13}$C ratio was assumed to be 60.  Furthermore, the ortho-to-para ratios were assumed to be 3 and 2 (statistical values) for H$_2$CO and D$_2$CO, respectively.  Although the ortho-to-para~ratio of H$_2$CO can be less than 3 in protostellar cores \citep[and references therein]{jor05b}, we ignored this effect for simplicity.  The derived column densities are summarized in Table \ref{tab3}.  The column densities are sensitive to the assumed excitation temperature.  A change in the excitation temperature by 5~K results in a change in the column densities by a factor of 2 - 3. 

In order to derive the beam-averaged column density of H$_2$, $N$(H$_2$), we employed the C$^{17}$O data.  The $N$(H$_2$) value was evaluated to be $(1.6\pm 0.3)\times 10^{23}$~cm$^{-2}$, $(1.0\pm 0.2)\times 10^{23}$~cm$^{-2}$, and $(8.0\pm 1.7)\times 10^{22}$~cm$^{-2}$, for $T=$ 15~K, 20~K, and 25~K, respectively, by assuming the $N$(C$^{17}$O)/$N$(H$_2$) ratio of $4.7 \times 10^{-8}$ \citep{fre82,cas99}.   
By using the derived $N$(H$_2$), the beam-averaged fractional abundances of molecules relative to H$_2$ ($X=N/N$(H$_2$)) were evaluated, as listed in Table \ref{tab4}.  The effect of the assumed excitation temperature is compensated in the fractional abundances.  The upper limits for some important molecules were also evaluated similarly by using the three times the r.m.s. noise at the expected frequency and the typical line width of 2~km~s$^{-1}$, as listed in Table \ref{tab4}.  

\subsection{Deuterium Fractionations} \label{deutf}
Spectral lines of the deuterated molecules DCO$^+$, DCN, CCD, HDCO, and D$_2$CO, were detected in this survey.  The column densities were derived under the assumption of LTE with $T_{\rm ex}$ = 20~K, as in the case of the normal (non-deuterated) species.  The column densities were also calculated for the excitation temperatures of 15~K and 25~K.  The results are included in Table \ref{tab3}.  The deuterium fractionation ratios are summarized in Table \ref{tab5}.  A change in the assumed excitation temperature by 5~K does not make a significant change in the ratio.  It should be noted that the deuterium fractionation ratio of D$_2$CO is almost comparable to that of HDCO.  

The DCO$^{+}$/HCO$^{+}$ ratio is lower than the deuterium fractionation ratios of the other species.  The ratio is consistent with that reported by \citet{and99} ($0.018 \pm 0.006$).  The DCO$^{+}$/HCO$^{+}$ ratio tends to reach its equilibrium ratio at the current temperature, since DCO$^{+}$ is readily destroyed by an electron recombination reaction.  On the other hand, the neutral species have longer lifetimes, and their deuterium fractionation ratios mostly remain as they were in the starless phase (Sakai et~al. 2011).  

\section{Discussion}
\subsection{Comparison with IRAS~16293-2422}
Our spectral line survey toward R~CrA~IRS7B is the second spectral line survey conducted for a low-mass star-forming region after the prototypical hot corino source IRAS~16293-2422 \citep{dis95,bla94,cau11}.  Hence, we compare the chemical composition of R~CrA~IRS7B with that of IRAS~16293-2422.  Figure \ref{fig5} shows a comparison of the fractional abundances of various molecules between R~CrA~IRS7B and IRAS~16293-2422.  The data for IRAS~16293-2422 are taken from \citet{bla94} and \citet{dis95}, where the abundances are derived under the same assumption as ours; constant $T_{\rm ex}$, no abundance variation within the source, and H$_{2}$ column density from the CO observation.  We can readily notice the following points:\\
(1)  In IRAS~16293-2422, the SO$_2$ lines are prominent, as in the case of hot cores in high-mass star-forming regions like Orion KL \citep[e.g.][]{bla86,sch97}.  On the other hand, the SO$_{2}$ lines are not as bright as in R~CrA~IRS7B.  This feature can be confirmed quantitively in the fractional abundance.  The abundance of SO$_2$ in R~CrA~IRS7B is lower by one order of magnitude than that in IRAS~16293-2422.  In contrast, the abundance of SO in R~CrA IRS7B is almost comparable to that in IRAS~16293-2422.  The other sulfur bearing organic species, CS, HCS$^{+}$, and H$_2$CS, also have similar abundances to the IRAS~16293-2422 case.  \\
(2)  The CH$_3$OH abundance is found to be lower by an order of magnitude than that in IRAS~16293-2422.  The rotation temperature of CH$_3$OH derived from the rotation diagram (31.0$\pm$6.8~K) is much lower than that reported for IRAS~16293-2422.  
The H$_2$CO abundance is, on the other hand, comparable to that in IRAS~16293-2422, although the rotation temperature of H$_2$CO (16.9$^{+5.4}_{-3.3}$~K) is lower than that of CH$_3$OH in R~CrA~IRS7B (Table \ref{tab2}).  \\
(3)  Complex organic molecules, such as HCOOCH$_3$, (CH$_3$)$_2$O, and C$_2$H$_5$CN, which are characteristic to hot corinos, are not detected.  This result does not directly mean that these species are deficient in R~CrA~IRS7B, since these molecules might be difficult to be excited in the 350~GHz region, as mentioned before.  Nevertheless, we can evaluate a meaningful upper limit for the (CH$_3$)$_2$O fractional abundance to be $4.7 \times 10^{-11}$ from the present observation.  This is significantly lower than the (CH$_3$)$_2$O fractional abundance in IRAS~16293-2422 \citep[$2.4 \times 10^{-7}$;][]{caz03}.  It is also lower than that in another hot corino, NGC 1333 IRAS2A \citep[$3.0 \times 10^{-8}$:][]{bot07}, and that in the high-mass star-forming region, Ori KL \citep[$8.0 \times 10^{-9}$:][]{bla86}.  As for the abundance of (CH$_3$)$_2$O relative to CH$_3$OH, we find the upper limit of 0.04 in R~CrA~IRS7B.  This upper limit is slightly lower than the corresponding ratio reported for IRAS~16293-2422 \citep[0.20:][]{herb09}.  It should be noted that these comparisons are based on the beam-averaged abundances.  For comparisons of the 'real' abundances, a detailed source model of R~CrR~IRS7B is necessary, which is left for future works.\\
(4)  The HCN abundance is comparable to that in IRAS~16293-2422, whereas the HNC abundance seems to be lower.  Hence, the HNC/HCN ratio is slightly lower than in the IRAS~16293-2422 case.  However, the ratio is within the range of ratios for starless cores and low-mass prestellar cores \citep{hiro98}.  In contrast, the CN abundance is much higher in R~CrA~IRS7B than that in IRAS~16293-2422.  This may indicate the importance of the photodissociation effect in R~CrA~IRS7B, as discussed below.  \\
(5) The CCH emission is very bright in R~CrA~IRS7B.  The CCH abundance is higher than that in IRAS~16293-2422 by more than one order of magnitude, although it is lower than that in the WCCC source L1527.  The relatively high abundance of CCH could be regarded as a sign of WCCC.  Alternatively, this may be another indication of the photodissociation effect.  The abundance of c-C$_3$H$_2$ in R~CrA~IRS7B is almost comparable to that in IRAS~16293-2422.  \\
(6)  The deuterium fractionation ratios of H$_2$CO are generally lower in R~CrA~IRS7B than in IRAS~16293-2422.  The HDCO/H$_2$CO ratio is $0.050 \pm 0.024$ in R~CrA~IRS7B, whereas it is 0.14 in IRAS~16293-2422 \citep{dis95}.  Although D$_2$CO is found in R~CrA~IRS7B and the D$_2$CO/H$_2$CO ratio is close to the HDCO/H$_2$CO ratio, these are general features for low-mass star-forming regions \citep{par06}.  Similarly, the CCD/CCH ratio is $0.038 \pm 0.016$ in R~CrA~IRS7B, which is lower than that in IRAS~16293-2422 (0.18).  In contrast, the DCN/HCN and DCO$^{+}$/HCO$^{+}$ ratios in R~CrA~IRS7B are similar to those in IRAS~16293-2422, and in the range of the ratios reported for low-mass protostellar sources \citep{jor04}.

\subsection{Origin of difference}
The chemical feature of R~CrA~IRS7B is completely different from that of IRAS~16293-2422, which is characterized by abundant complex organic molecules and SO$_2$.  The results (1), (2), (3), and (6) suggest that the hot corino activity is weaker than previously thought in R~CrA~IRS7B.  It seems that the bright H$_2$CO and CH$_3$OH emission does not directly represent strong hot corino activities.  On the other hand, molecules related to carbon-chain molecules such as CCH and c-C$_3$H$_2$ are relatively abundant, as mentioned in (5), and the deuterium fractionation ratios are relatively low (6).  These features are generally seen in WCCC sources \citep{sak09b}.  However, longer carbon-chain molecules such as C$_4$H and HC$_3$N are not detected in the 345~GHz band.  Furthermore, the C$_4$H line is not detected in the 3 mm observations using Mopra \citep{sak09a}.  Hence, R~CrA~IRS7B cannot definitively be categorized to the WCCC sources in the present stage.  

One possibility of the distinct chemical composition from the two categories is that R~CrA~IRS7B has an intermediate characteristic between them.  \citet{sak09a} proposed that the difference between hot corinos and WCCC sources originates from different chemical compositions of the grain mantles caused by different duration time of the starless core phase.  According to their scenario, the hot corino and WCCC sources are the two extremes, and the existence of sources with a mixture of these chemical characteristics is plausible.  However, it is difficult to confirm or disprove this picture from the present observations, since the emission lines of long carbon-chain molecules and large complex organic molecules are difficult to be excited in the submillimeter-wave region.  Rather this picture could be tested by observations in lower frequency bands.   

Given the proximity of the nearby Herbig Ae star R~CrA located outside of the protostar R~CrA~IRS7B, the chemical composition of the envelope of R~CrA~IRS7B would be affected by the external UV radiation.  \citet{sta07} discussed the effect of the FUV and X-ray radiation from protostars, based on their observation of CN, CO$^{+}$, SO, SO$^{+}$, NO, and HCN.  According to their result, the CN/HCN, CN/NO, SO$^{+}$/SO, and CO$^{+}$/HCO$^{+}$ ratios are higher in the envelope of high-mass stars compared to those surrounding in low-mass stars.  This trend is interpreted in terms of the photodissociation effect due to the bright FUV radiation from high-mass young stellar objects on their ambient envelopes.  Comparing our results with theirs, we find that the abundance ratios of CN/HCN, CN/NO, and SO$^{+}$/SO are evaluated to be 1.8, 0.7, and 0.016, respectively, which are comparable to the ratios of high-mass star-forming regions as  by \citet{sta07} (e.g 1.80, 0.21, and 0.007, respectively for W3 IRS5).  On the other hand, these ratios are an order of magnitude higher in R~CrA~IRS7B than in IRAS~16293-2422.  Therefore, the envelope of R~CrA~IRS7B could be affected by photodissociation.  In this case, the source of the UV radiation is not the protostar within R~CrA~IRS7B but nearby Herbig Ae star R~CrA which is situated at 39 arcsec NW of R~CrA~IRS7B.  Note that the C/CO ratio toward R CrA IRS7B is evaluated to be 0.06 - 0.09 by assuming the CO/H$_2$ ratio of $10^{-4}$.  This ratio is consistent with that found in the $\rho$ Ophiuchi cloud illuminated by nearby B star (HD147889) \citep{kam03}.  Furthermore, the 4.62 $\mu$m "XCN" absorption feature is marginally detected toward R~CrA~IRS7B \citep{chi98, whi01}.  Formation of "XCN" requires energetic phenomena such as UV photolysis \citep{gri87} and ion bombardment \citep{pal00}.  This might also support the PDR picture mentioned above.

As mentioned in Section 3.3, the deuterium fractionation ratios of the neutral species reflect the initial conditions in the prestellar cores before the onset of star formation.  The low deuterium fractionation ratios may suggest that the parent core of R~CrA~IRS7B was so warm in the starless core phase that the CO molecule had survived against depletion.  If the parent core was heated by the external UV radiation from R~CrA, the low deuterium fractionation ratio could also be explained.  

Although the above results are consistent with predictions for PDRs, the other abundance ratios, which are also characteristic in PDR, show inconsistency.  The CO$^{+}$/HCO$^{+}$ ratio is found to be $<$ 0.002, which is much lower than those in high-mass star-forming regions (0.016 - 0.066), and similar to that in hot corinos (0.002 - 0.001) \citep{sta07}.  The HCO$^{+}$/HOC$^{+}$ ratio is evaluated to be $>$ 714, which is higher than those in PDRs reported by \citet{fue03} (50 - 120 for the PDR in NGC 7023).  Moreover, the UV source is located outside the protostellar envelope of R~CrA~IRS7B, in contrast to the high-mass star-forming regions where the UV sources are embedded inside the envelope.  Therefore, the effect of FUV and X-ray on the chemical composition in R~CrA~IRS7B could be different from the high-mass star-forming region case.  For a further understanding of those effects, we need more detailed analyses, including a source model and chemical network studies.  

\section{Summary}
\begin{enumerate}
\item We have carried out a spectral line survey toward Class 0 protostar R~CrA~IRS7B in Corona Australis dark cloud in the 345~GHz band with ASTE, as part of our multi-wavelength spectral line survey project of R~CrA~IRS7B.  In this survey, 16 molecular species and 16 isotopomers are identified.  We have also made a supplementary observation in the 450~GHz band, where 3 molecular and one atomic species are detected.
\item We have found very bright CN and CCH emissions, whereas the SO$_2$ lines are not as prominent as in IRAS~16293-2422.  The abundances of CN and CCH are higher than those in IRAS~16293-2422 by an order of magnitude.  In contrast, the abundances of SO$_2$ and CH$_3$OH are much lower than those in IRAS~16293-2422.  
\item In the 345~GHz band, neither complex organic molecules nor long charbon-chain molecules, which consist of 4 heavy atoms or more, are detected.  However, this does not directly mean that these species are deficient in R~CrA~IRS7B, since their emission lines in the submillimeter-wave region are difficult to be excited.  
\item Deuterium fractionation ratios are obtained for CCH, H$_2$CO, HCN, and HCO$^{+}$.  They are less than 5~\%.  The HDCO/H$_2$CO and CCD/CCH ratios are much lower than those reported for IRAS~16293-2422.  
\item Excitation temperatures of SO$_2$, CH$_3$OH, c-C$_3$H$_2$, H$_2$CO, and CCH are all similar to one another, indicating that these molecules reside in the same region.  Moreover, the excitation temperature of CH$_3$OH is much lower than that found in IRAS~16293-2422.  
\item From these results, we find that the chemical composition of R~CrA~IRS7B is clearly different from that of hot corinos.  However, R~CrA~IRS7B cannot be categorized to the WCCC sources, since long carbon-chain molecules, which are commonly found in WCCC source, are not detected.  If the hot corinos and the WCCC sources are the two extremes with respect to chemical compositions of Class 0 objects, R~CrA~IRS7B would be a source with a mixture of these two chemical characteristics.
\item Alternatively, the chemical composition of R~CrA~IRS7B would be significantly affected by the UV radiation from the nearby Herbig Ae star R~CrA, as indicated by the bright CN emission.  The effect of the UV radiation has to be examined by detailed chemical models, which is left for future works.  
\item A spectral line survey is a powerful technique to characterize the chemical features of the protostellar sources.  The present survey indicates further chemical diversity in low-mass star-forming regions.  Similar surveys toward the other low-mass star-forming regions are of particular importance to explore the origin of their chemical diversity.  
\end{enumerate}

\section{Acknowledgements}
The authors are grateful to the ASTE staff for excellent support.  The research of JKJ is supported by a Junior Group Leader Fellowship from the Lundbeck foundation.  The research in Copenhagen (JL, JKJ and SEB) is furthermore supported by a grant from Instrumentcenter for Danish Astrophysics and by Centre for Star and Planet Formation, which is funded by the Danish National Research Foundation and the University of Copenhagen's programme of excellence for financial support.  YW acknowledges to the Hayakawa Satio Fund awarded by the Astronomical Society of Japan.  This study is supported by a Grant in Aid from the Ministory of Education, Culture, Sports, Science, and Technology of Japan (No. 21224002 and 21740132).

\clearpage

%%  Figures

\clearpage
\begin{figure}
\epsscale{1.05}
\plotone{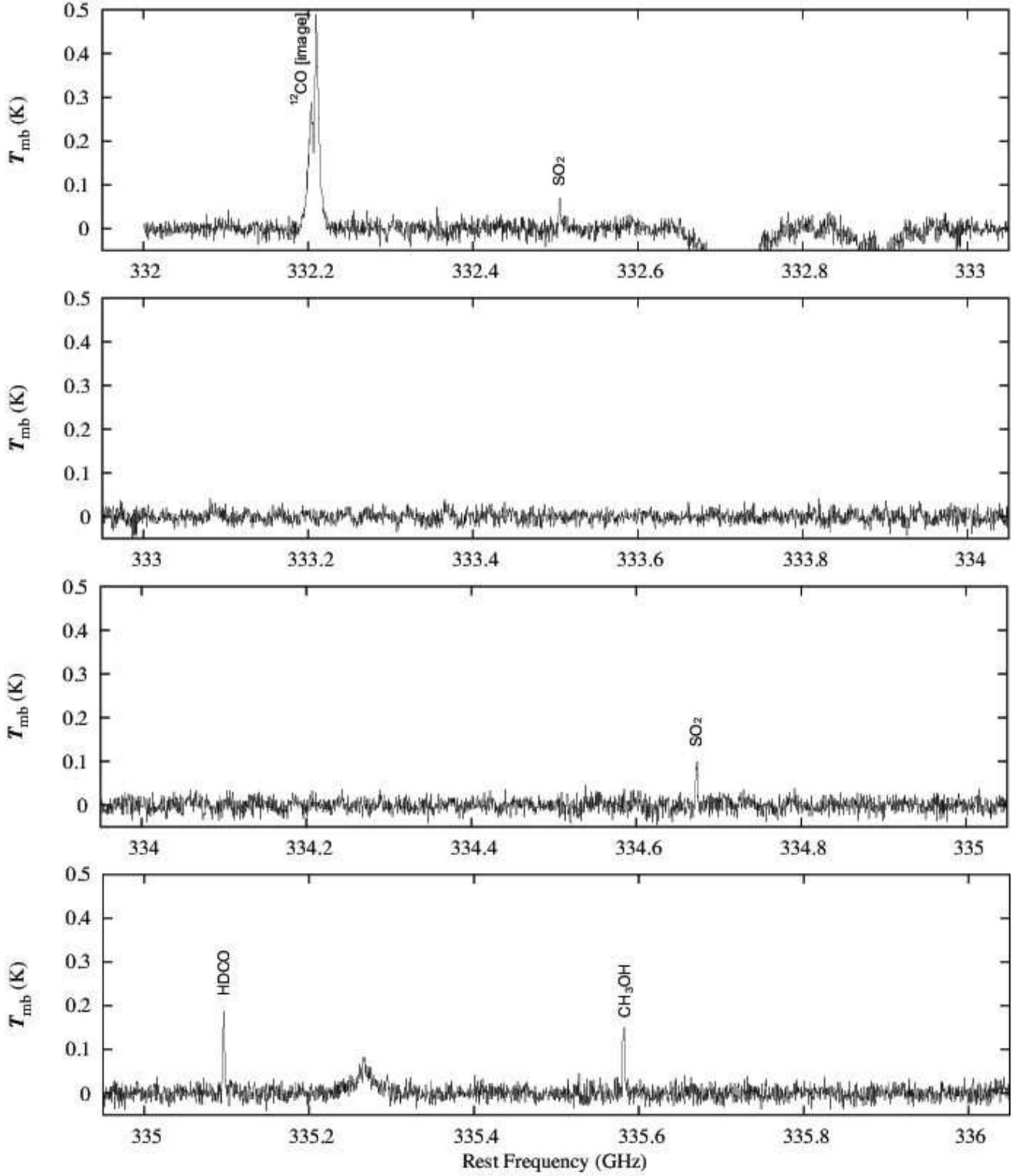}
\caption{Spectrum of R~CrA~IRS7B.  \label{fig2}}
\end{figure}
\setcounter{figure}{1}

\clearpage
\begin{figure}
\epsscale{1.05}
\plotone{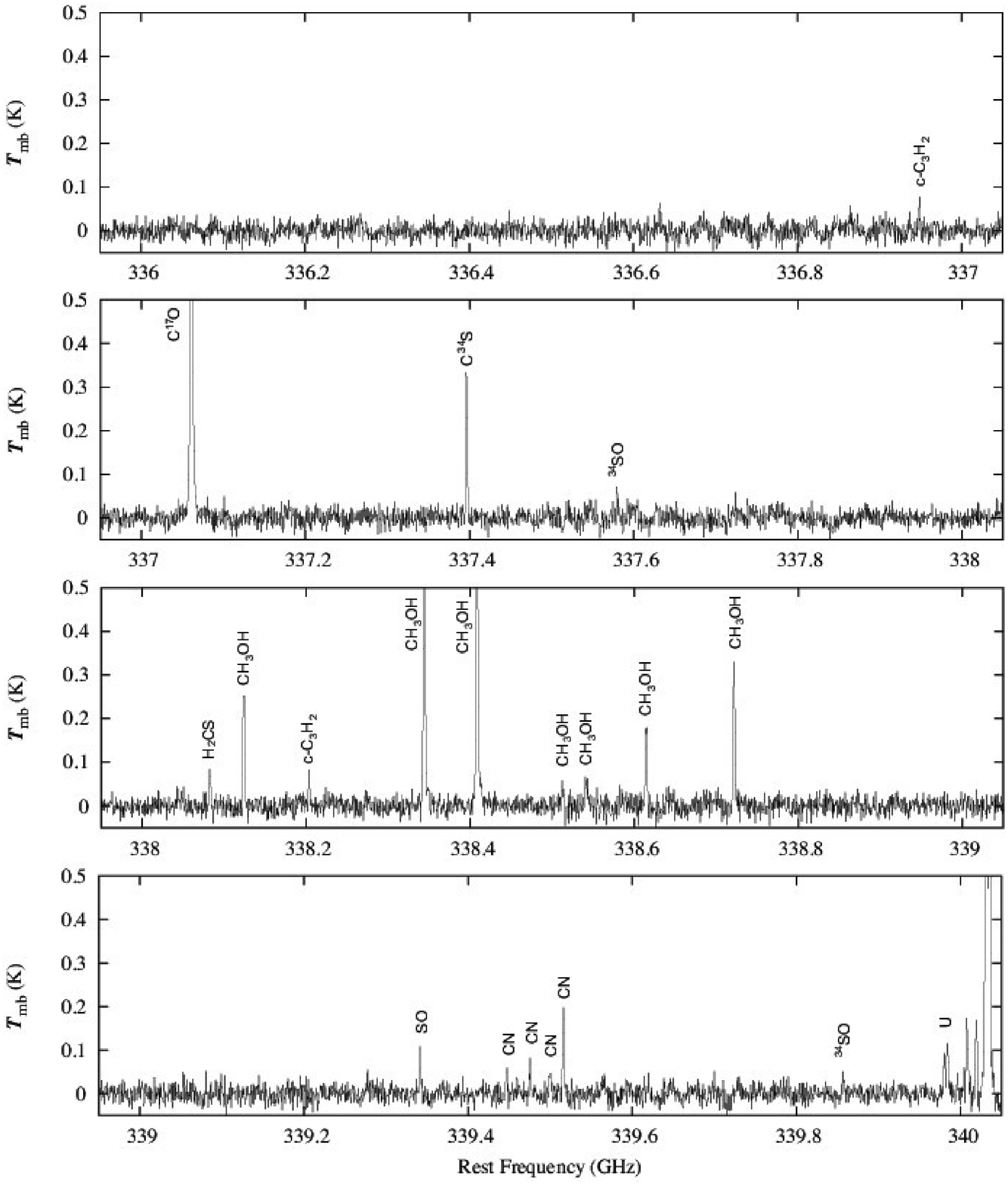}
\caption{\textit{Continued}}
\end{figure}
\setcounter{figure}{1}

\clearpage
\begin{figure}
\plotone{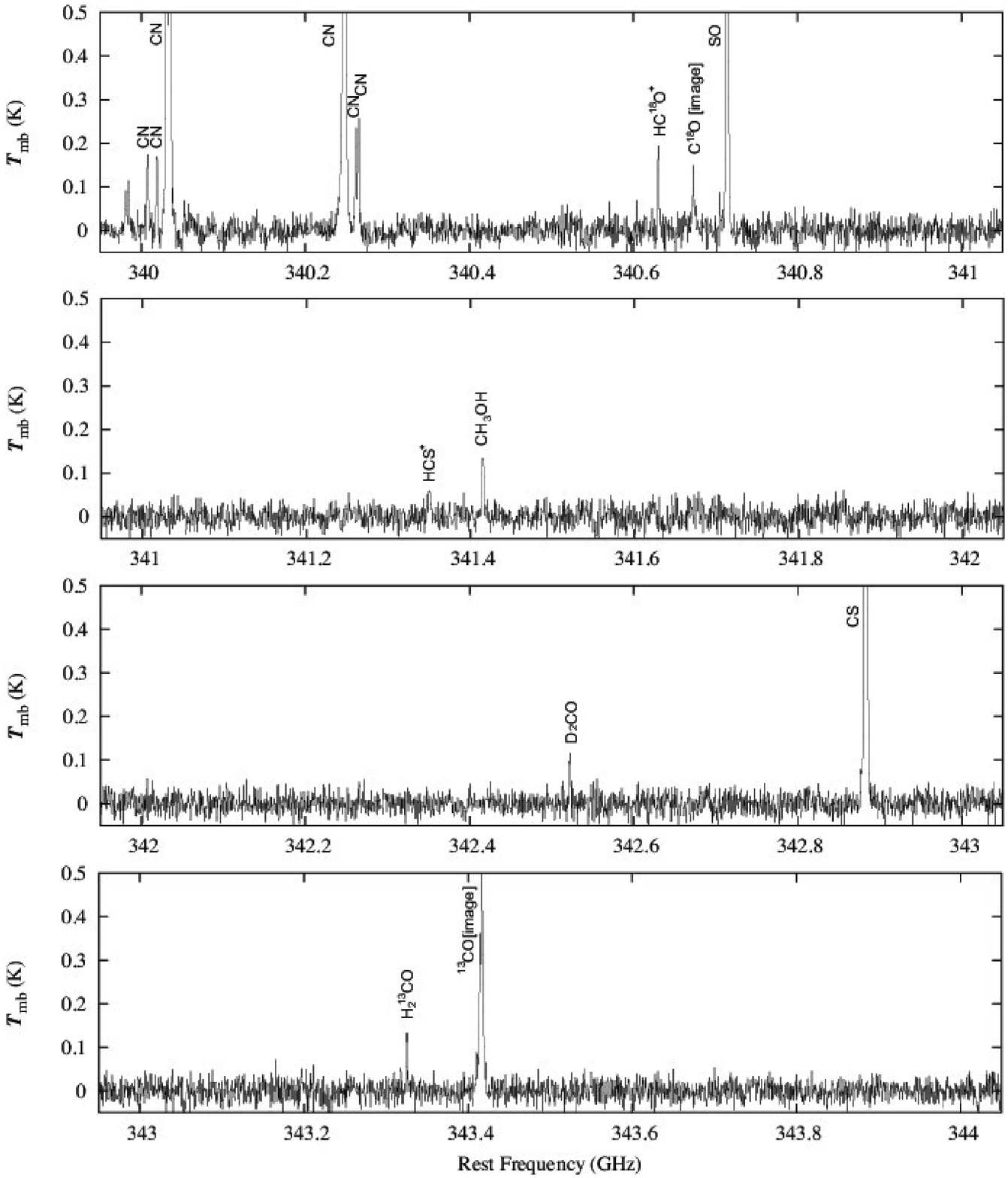}
\caption{\textit{Continued}}
\end{figure}
\setcounter{figure}{1}

\clearpage
\begin{figure}
\plotone{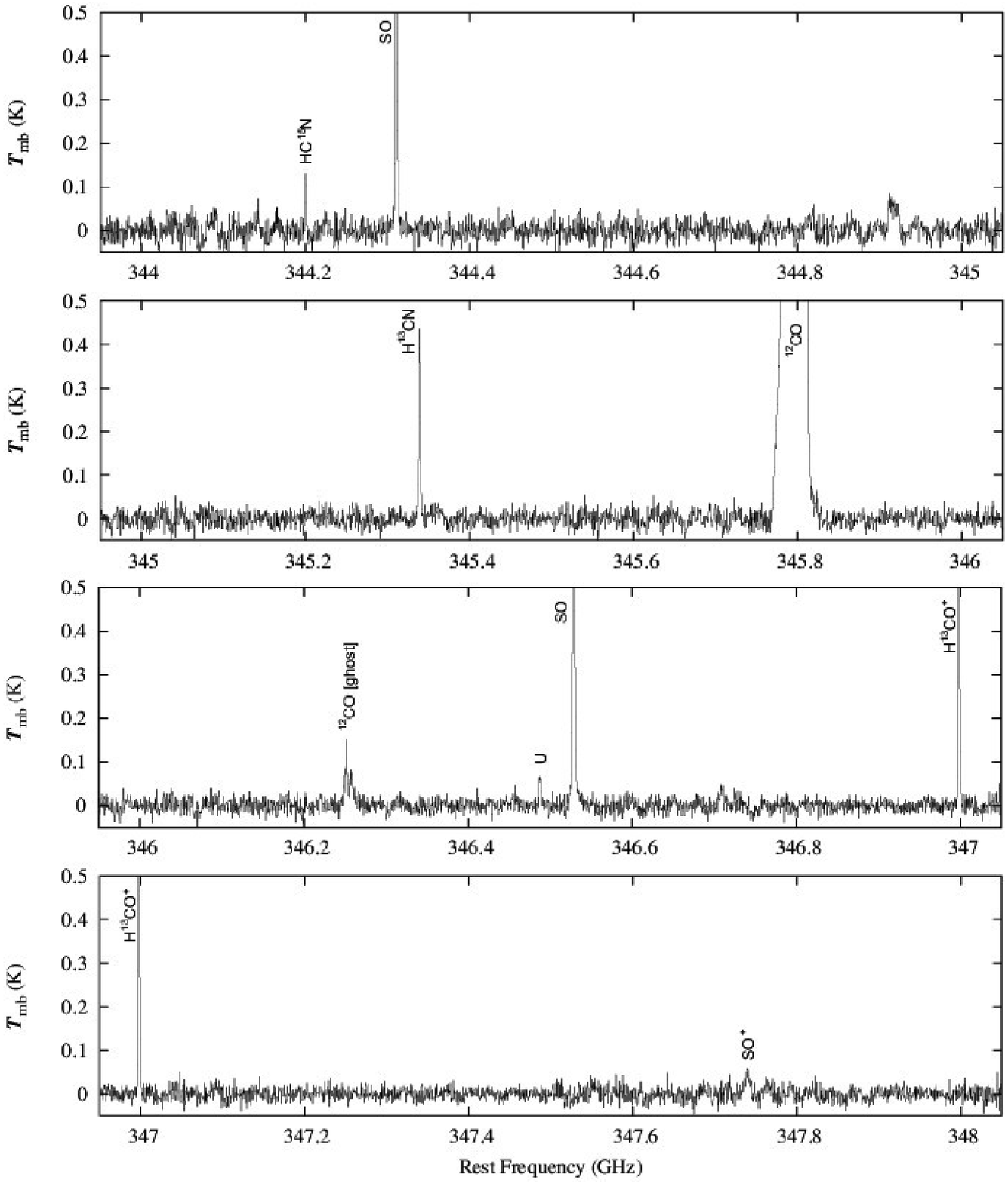}
\caption{\textit{Continued}}
\end{figure}
\setcounter{figure}{1}

\clearpage
\begin{figure}
\plotone{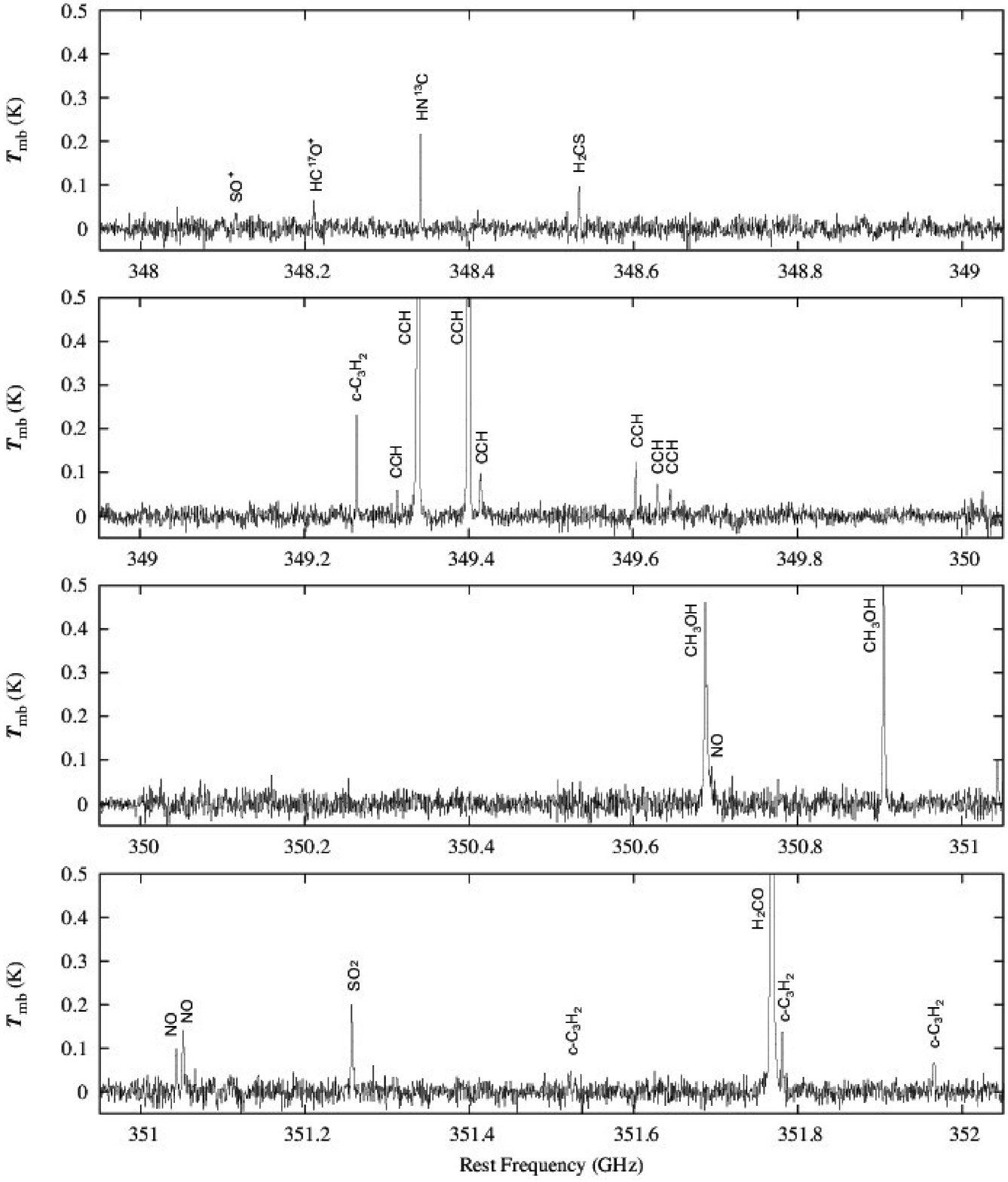}
\caption{\textit{Continued}}
\end{figure}
\setcounter{figure}{1}

\clearpage
\begin{figure}
\plotone{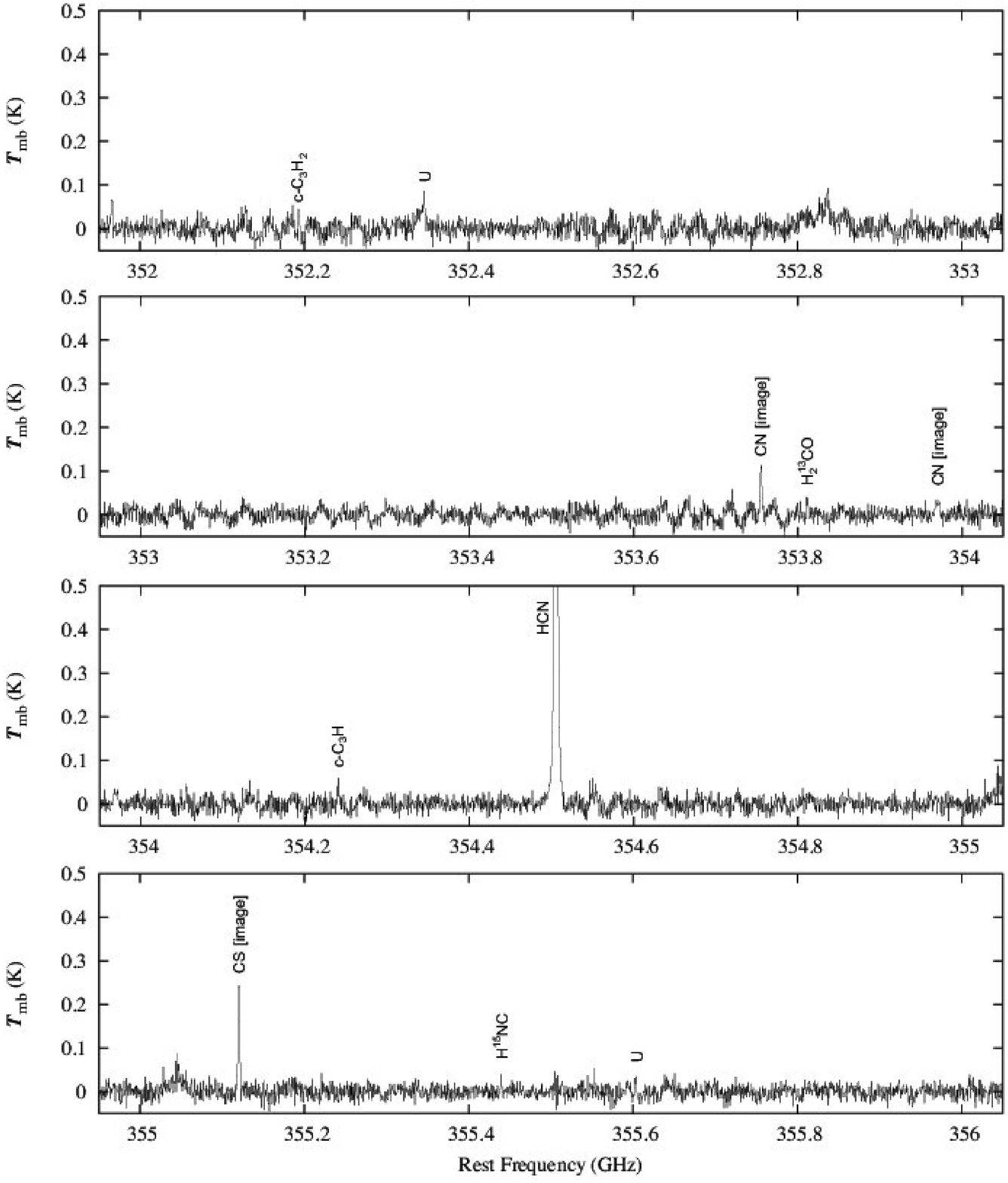}
\caption{\textit{Continued}}
\end{figure}
\setcounter{figure}{1}

\clearpage
\begin{figure}
\plotone{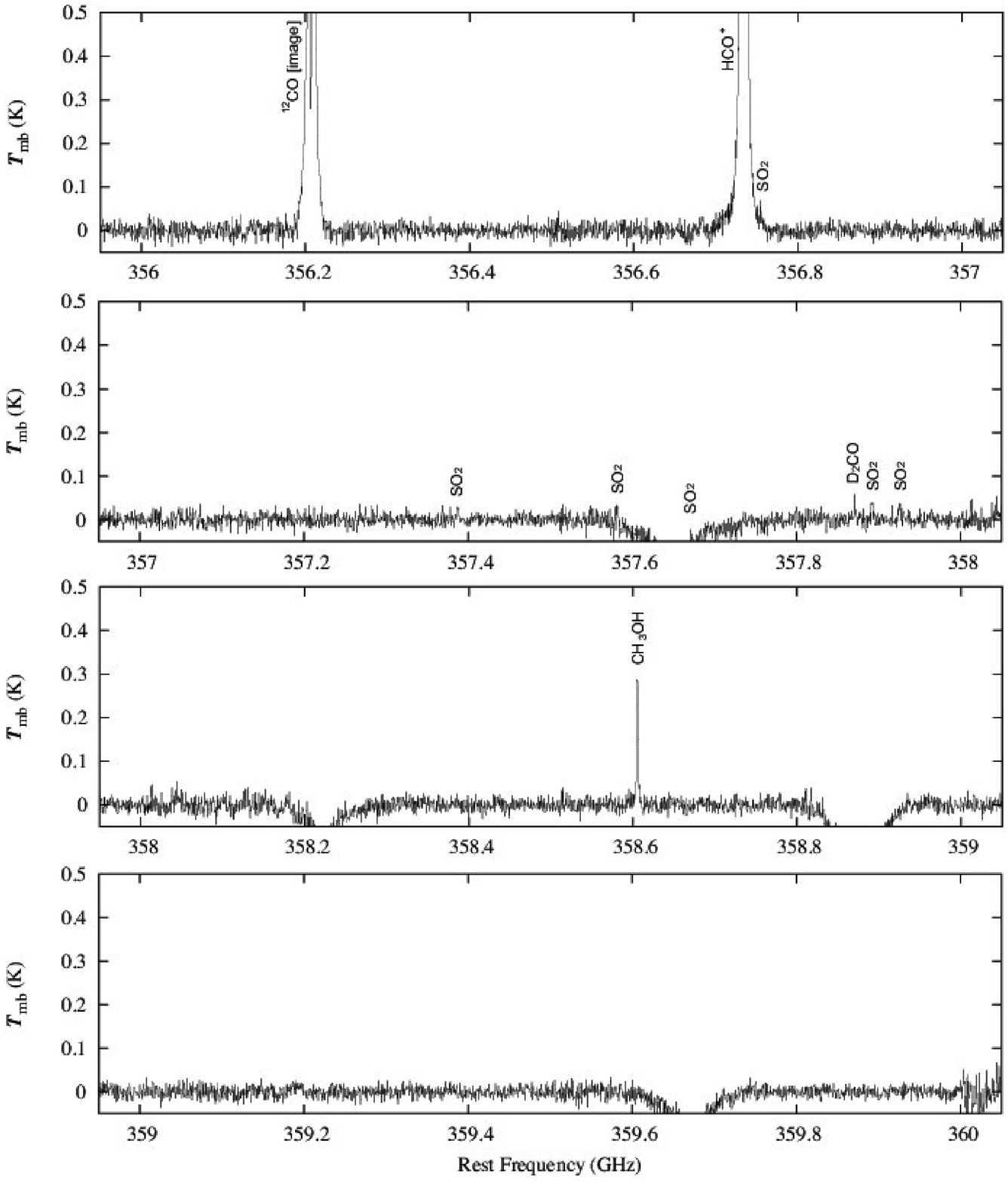}
\caption{\textit{Continued}}
\end{figure}
\setcounter{figure}{1}

\clearpage
\begin{figure}
\plotone{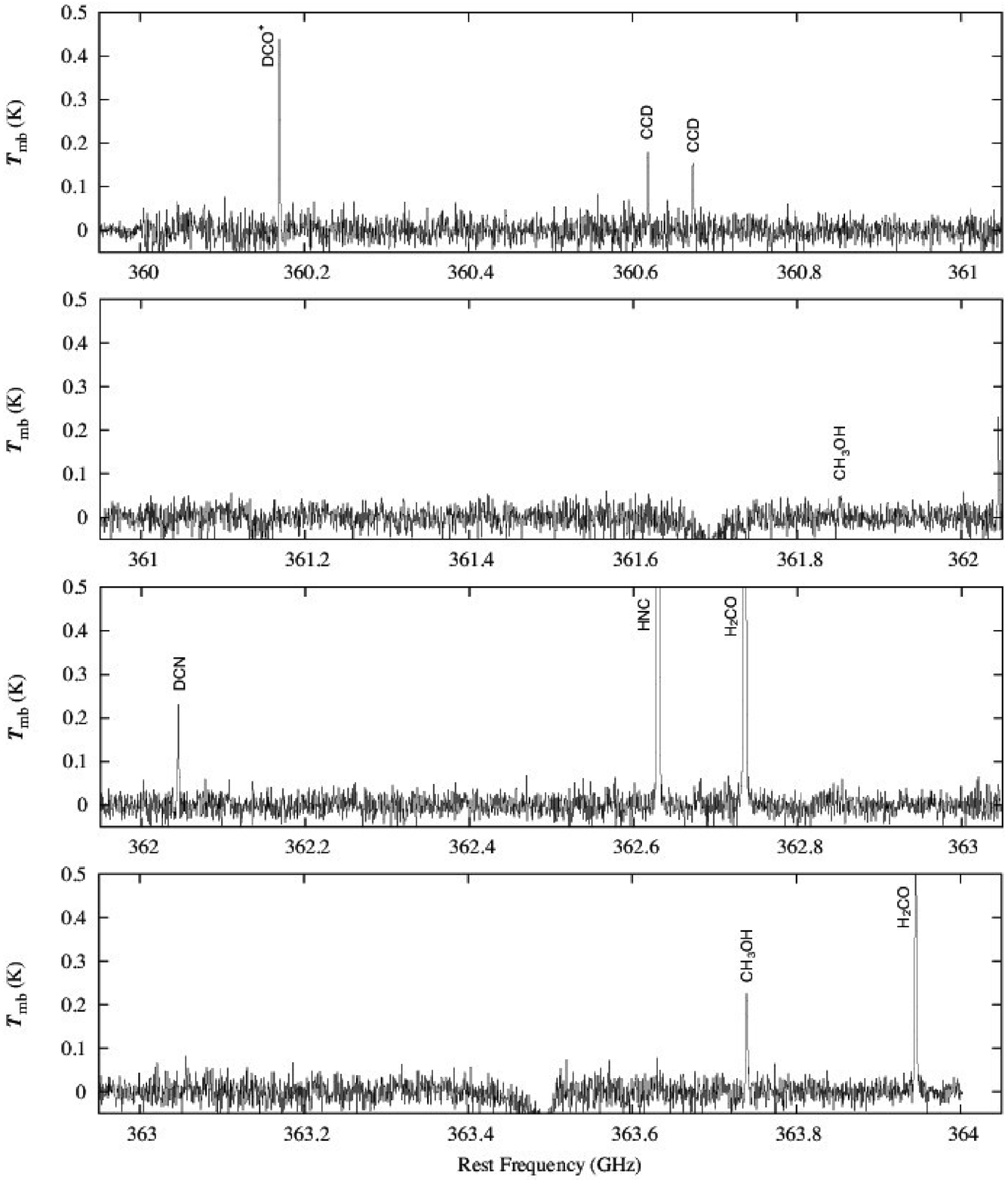}
\caption{\textit{Continued}}
\end{figure}
\setcounter{figure}{1}

\clearpage
\begin{figure}
\plotone{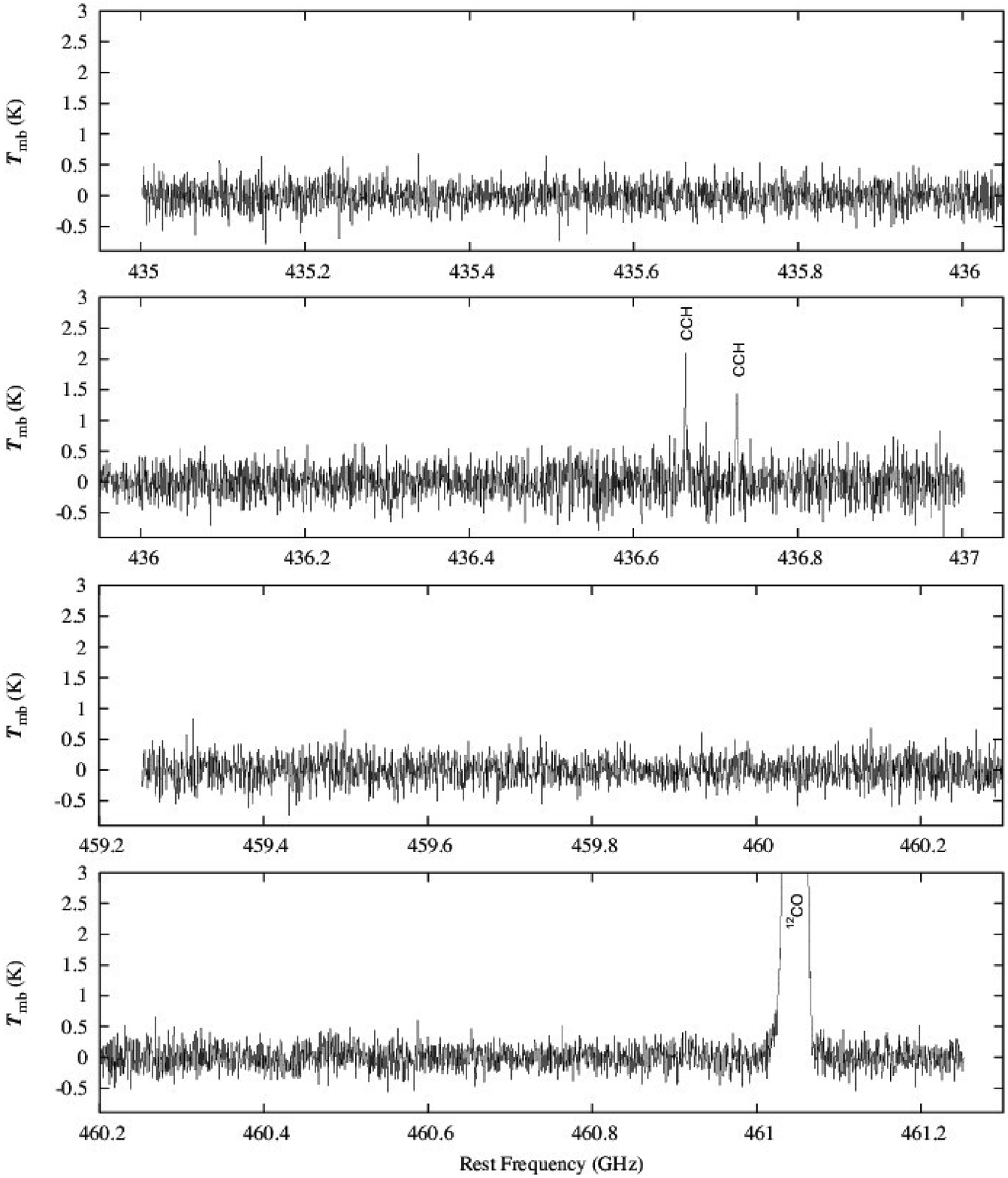}
\caption{\textit{Continued}}
\end{figure}
\setcounter{figure}{1}

\clearpage
\begin{figure}
\plotone{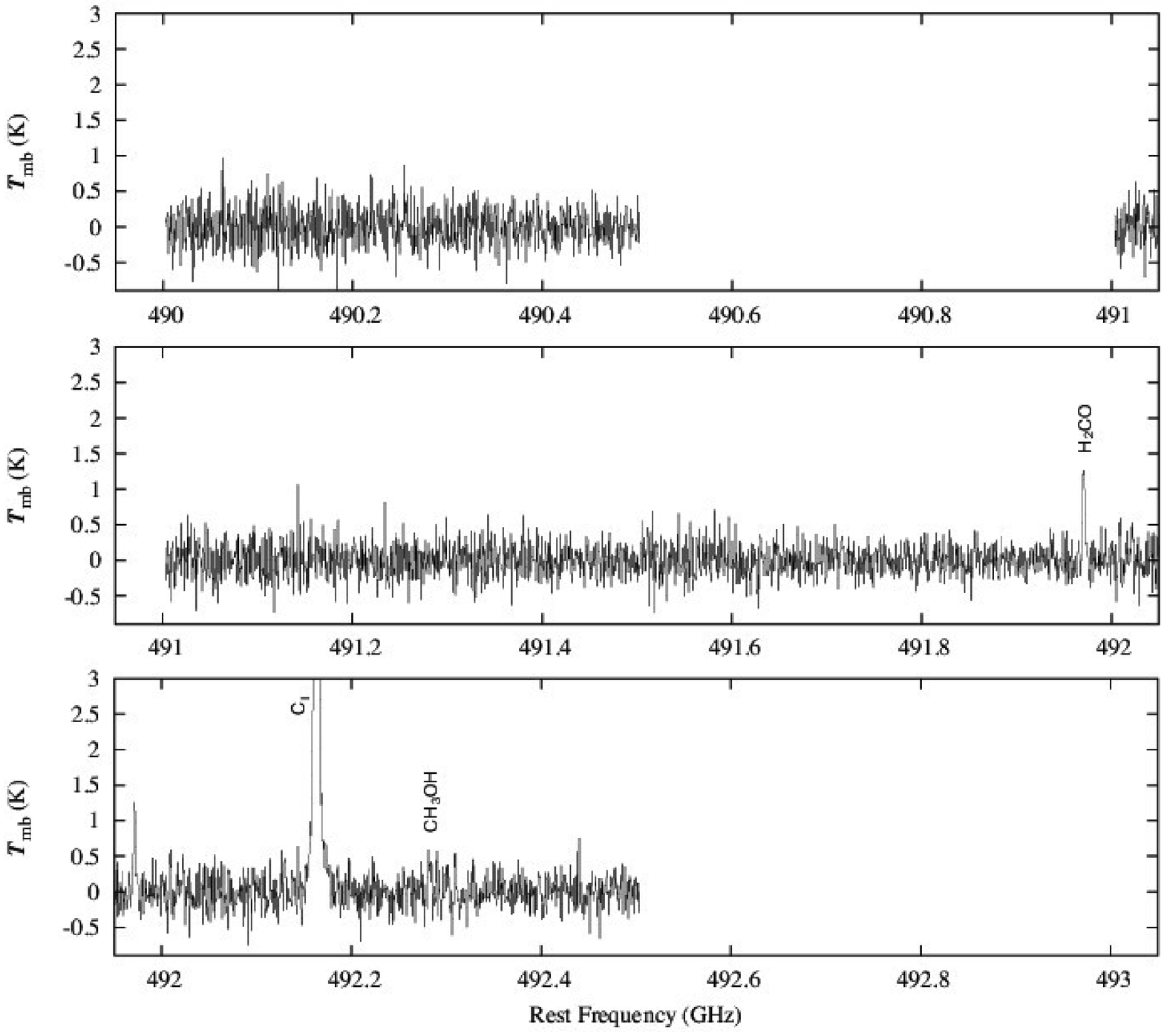}
\caption{\textit{Continued}}
\end{figure}

\clearpage
\begin{figure}
\includegraphics[angle=90, scale=0.53]{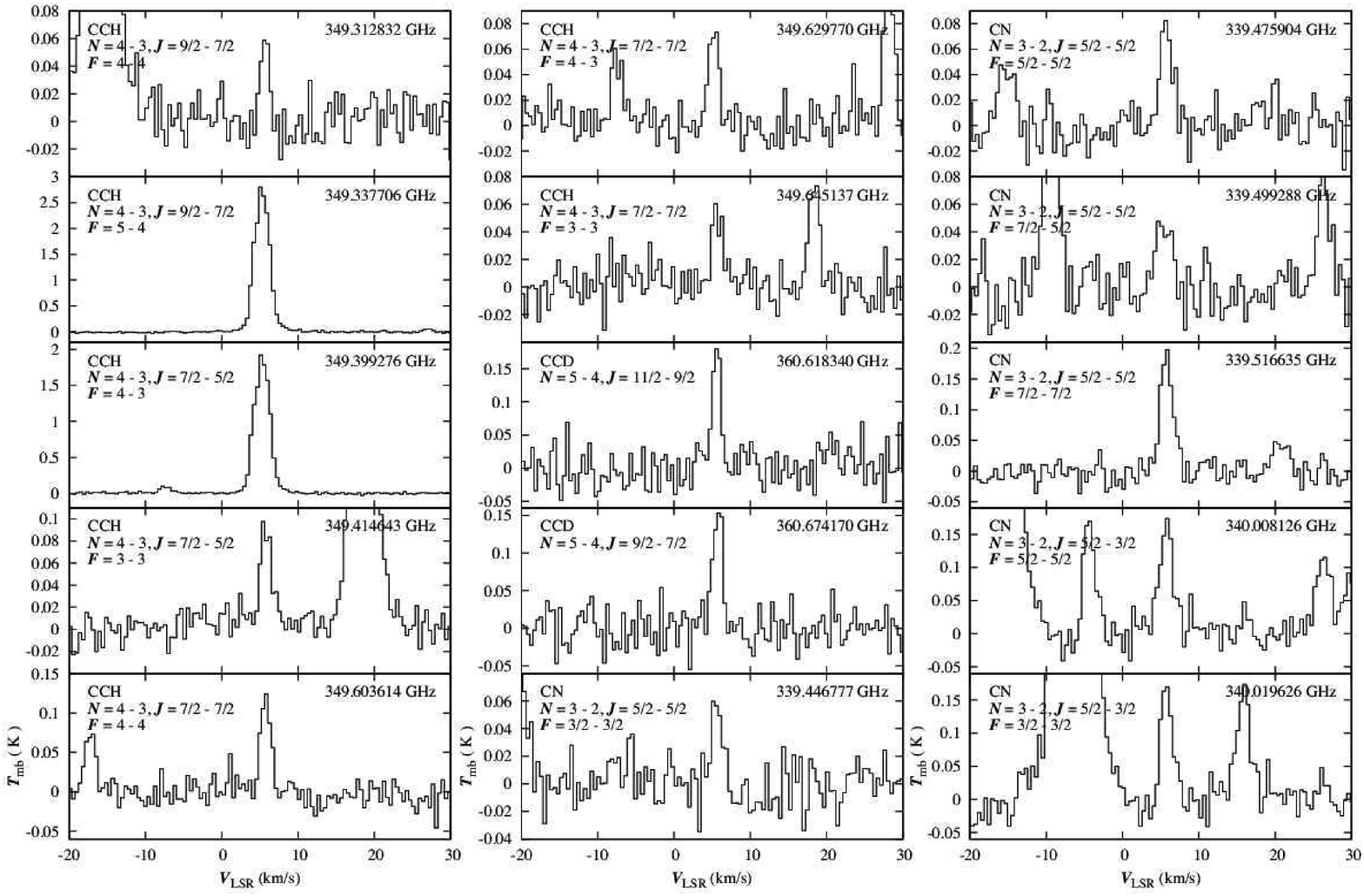}
\caption{Spectra of individual molecules observed in R~CrA~IRS7B.  \label{fig3}}
\end{figure}
\setcounter{figure}{2}

\clearpage
\begin{figure}
\includegraphics[angle=90, scale=0.53]{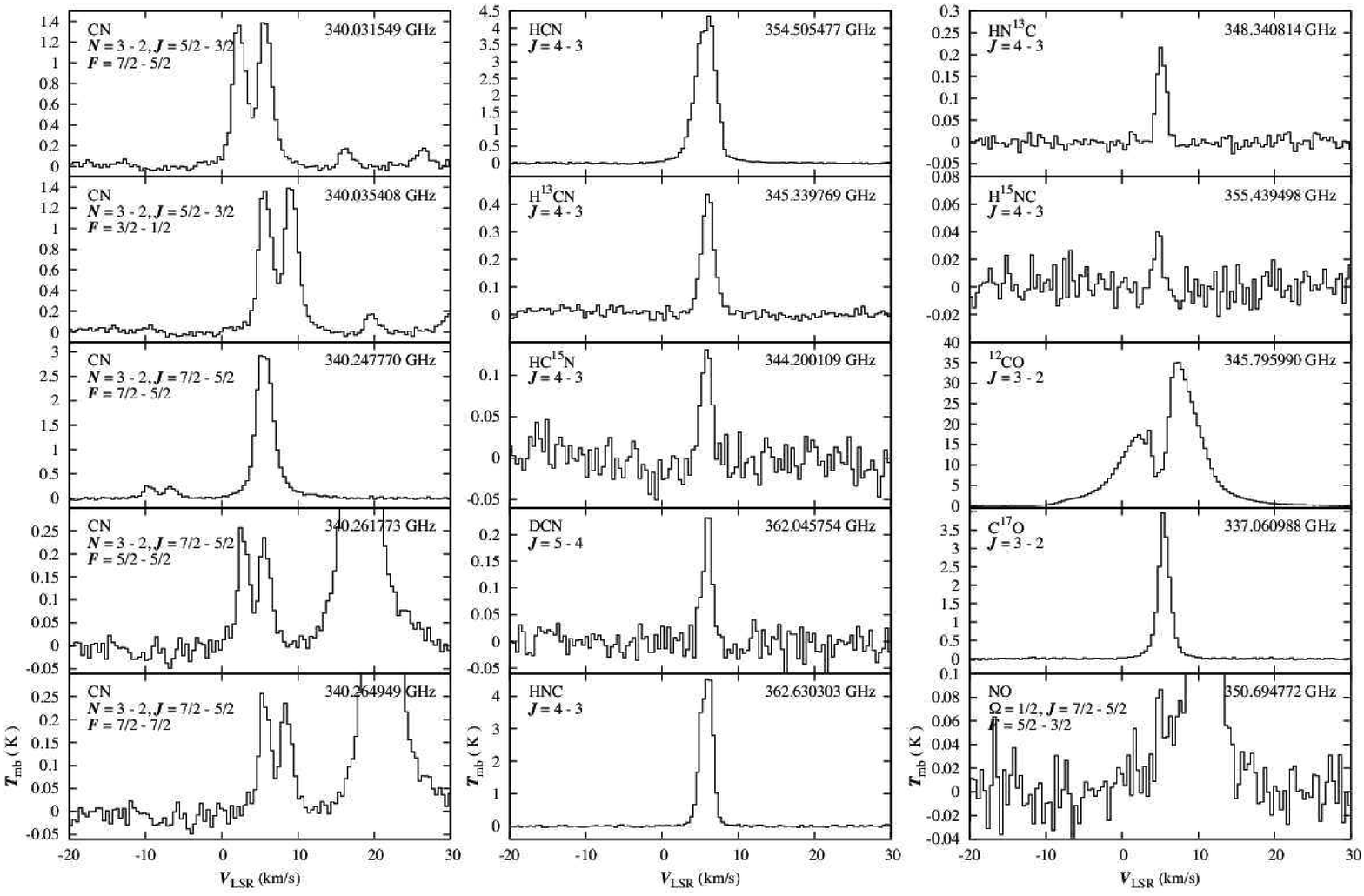}
\caption{\textit{Continued}}
\end{figure}
\setcounter{figure}{2}

\clearpage
\begin{figure}
\includegraphics[angle=90, scale=0.53]{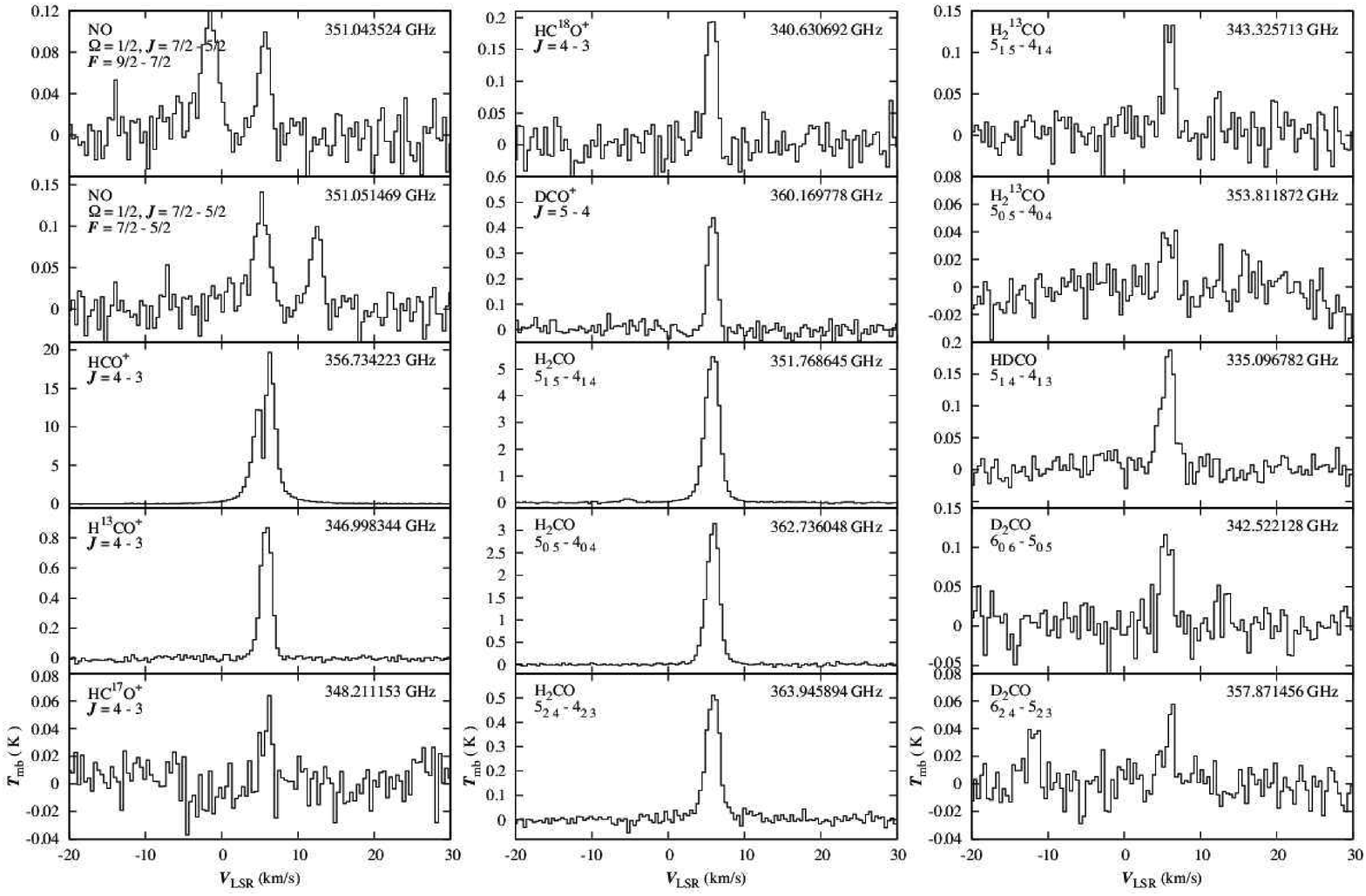}
\caption{\textit{Continued}}
\end{figure}
\setcounter{figure}{2}

\clearpage
\begin{figure}
\includegraphics[angle=90, scale=0.53]{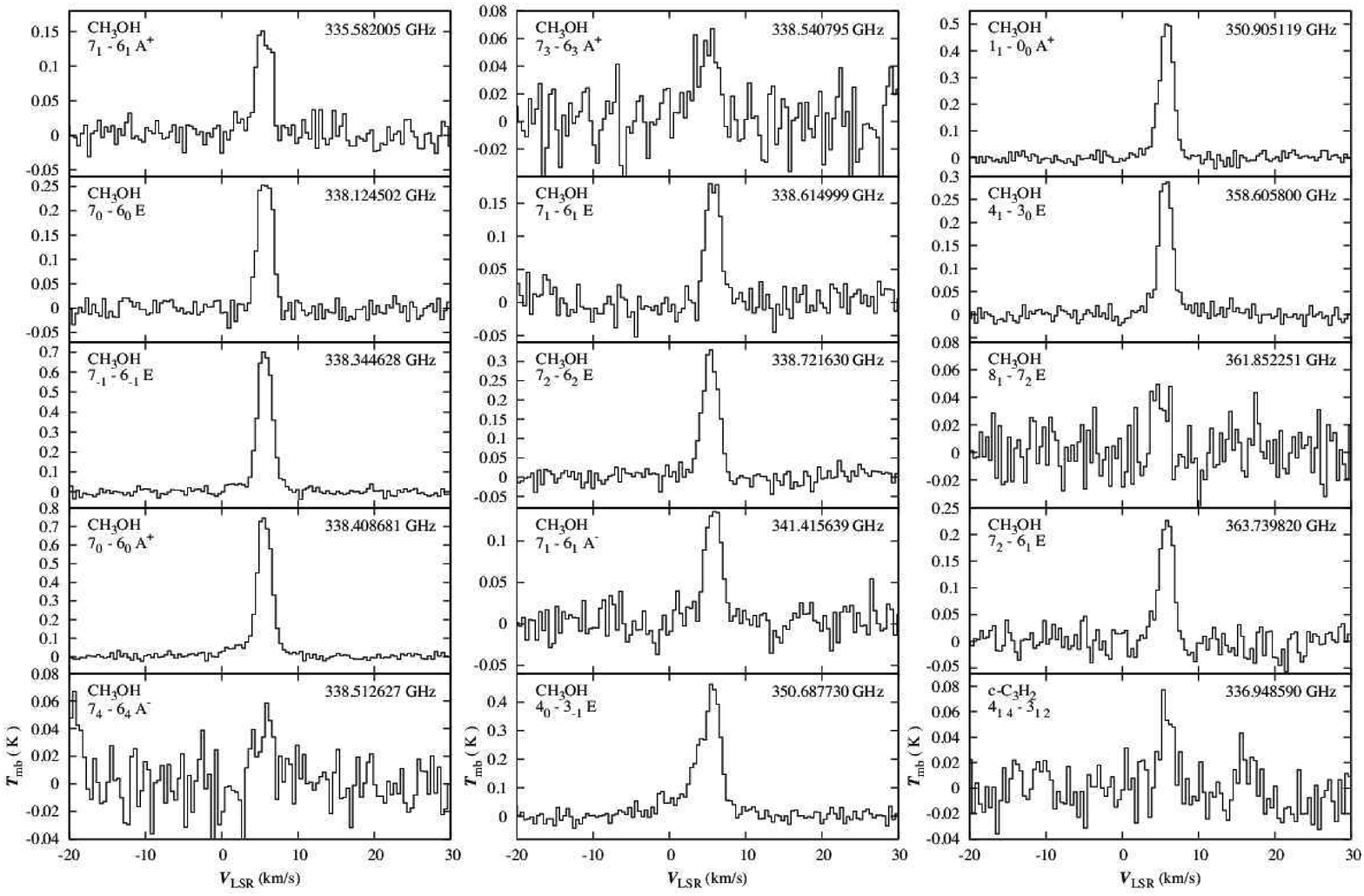}
\caption{\textit{Continued}}
\end{figure}
\setcounter{figure}{2}

\clearpage
\begin{figure}
\includegraphics[angle=90, scale=0.53]{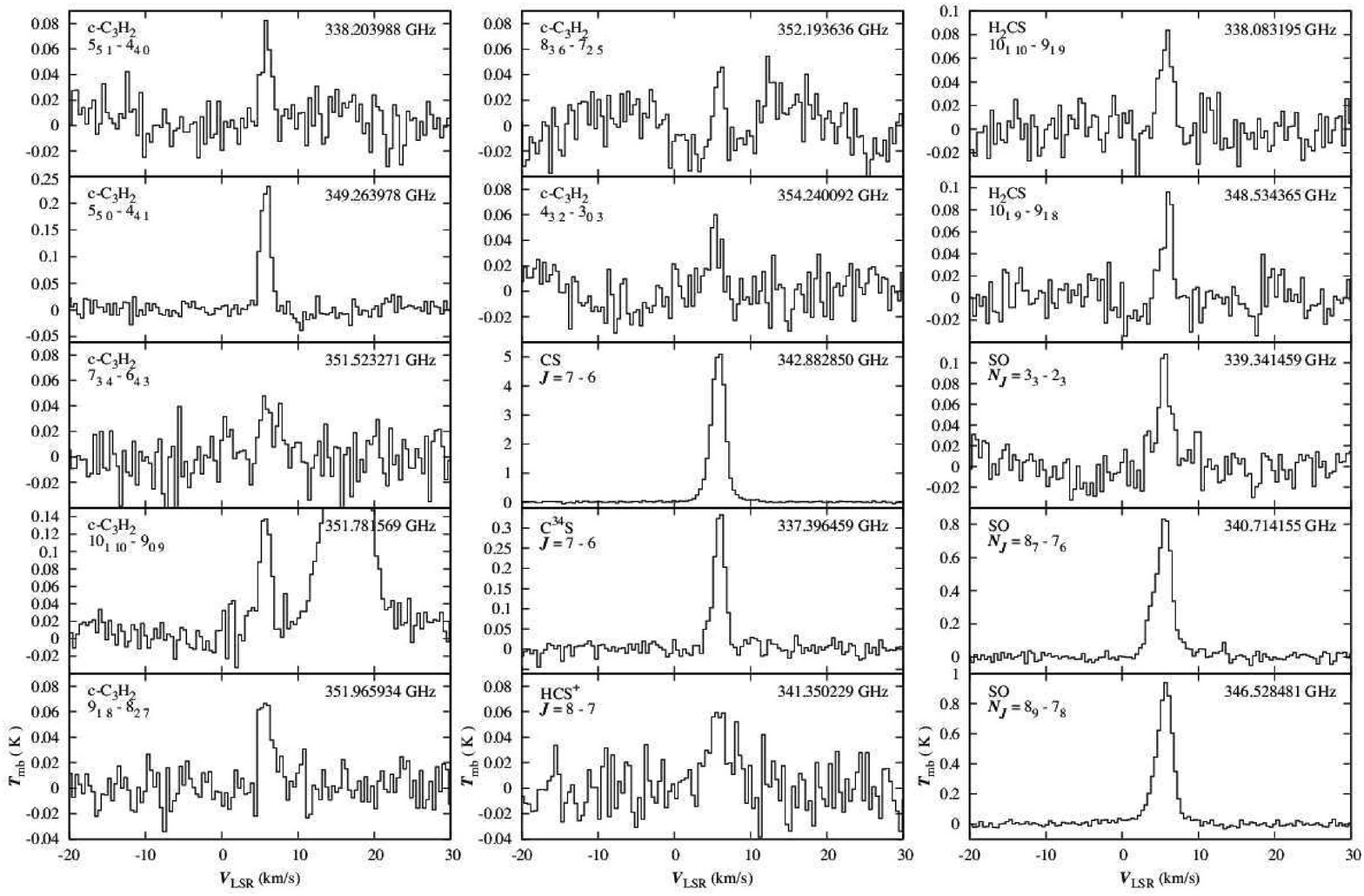}
\caption{\textit{Continued}}
\end{figure}
\setcounter{figure}{2}

\clearpage
\begin{figure}
\includegraphics[angle=90, scale=0.53]{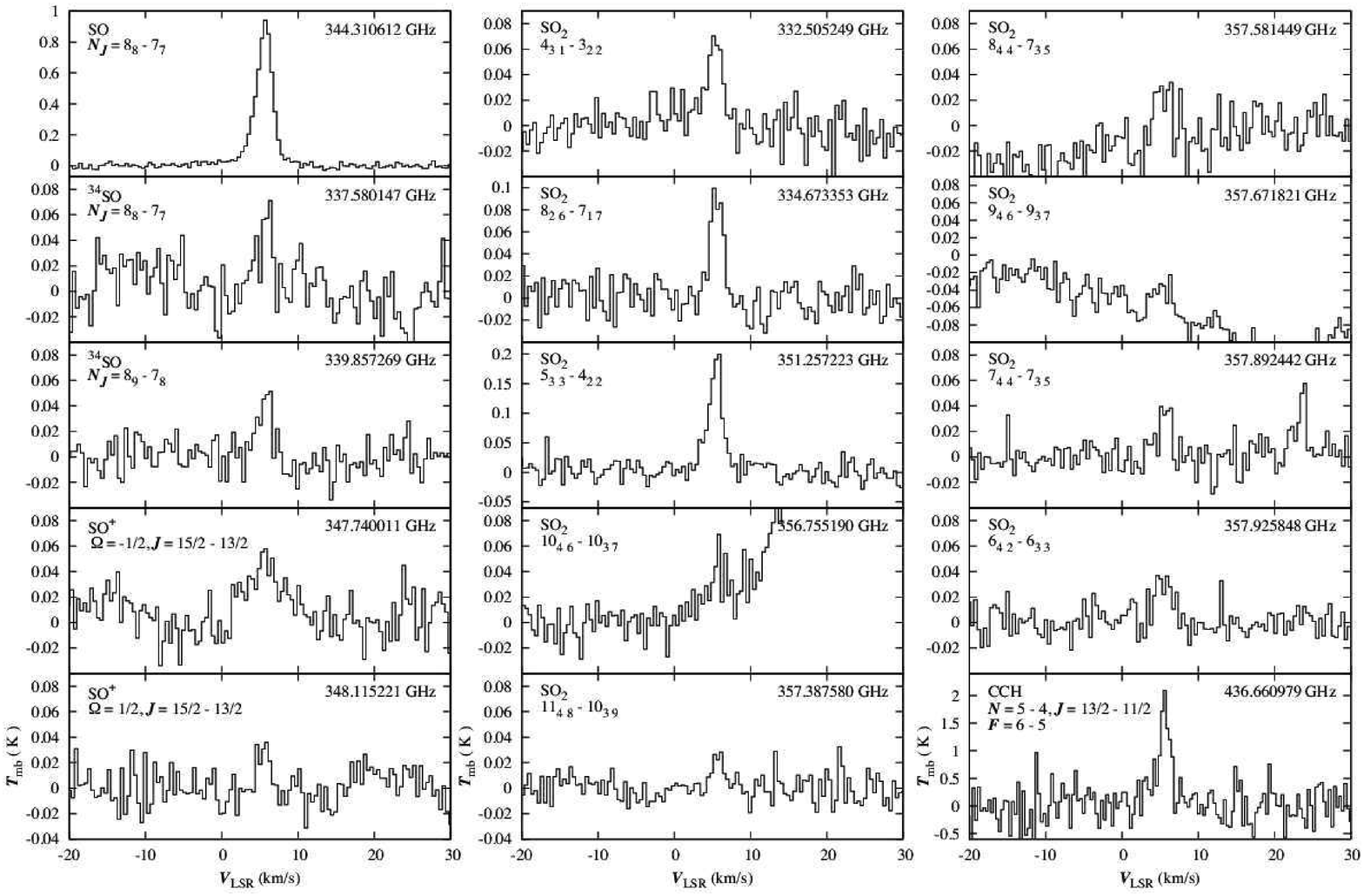}
\caption{\textit{Continued}}
\end{figure}
\setcounter{figure}{2}

\clearpage
\begin{figure}
\includegraphics[angle=90, scale=0.53]{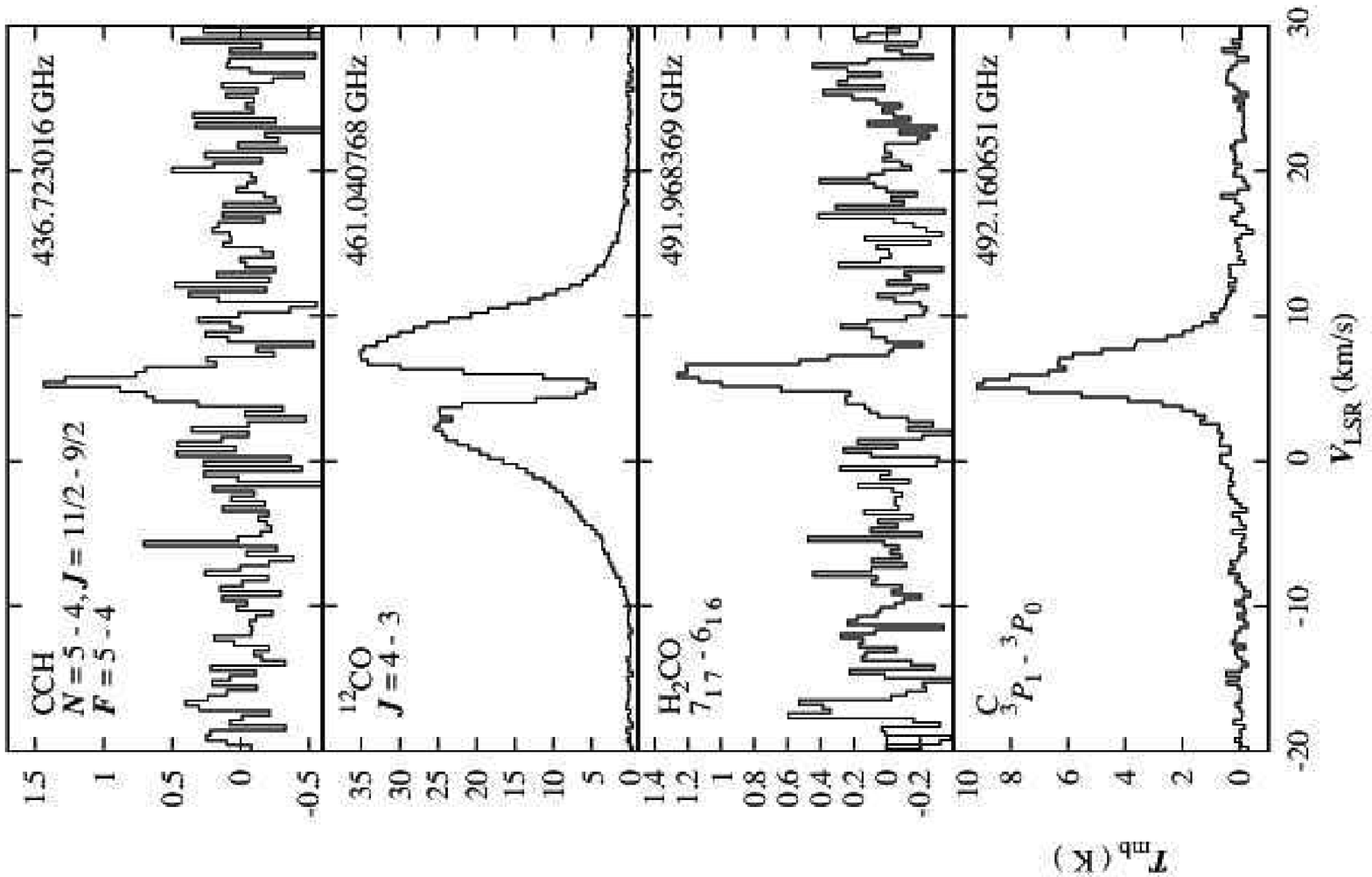}
\caption{\textit{Continued}}
\end{figure}

\clearpage
\begin{figure}
\epsscale{1.0}
\plotone{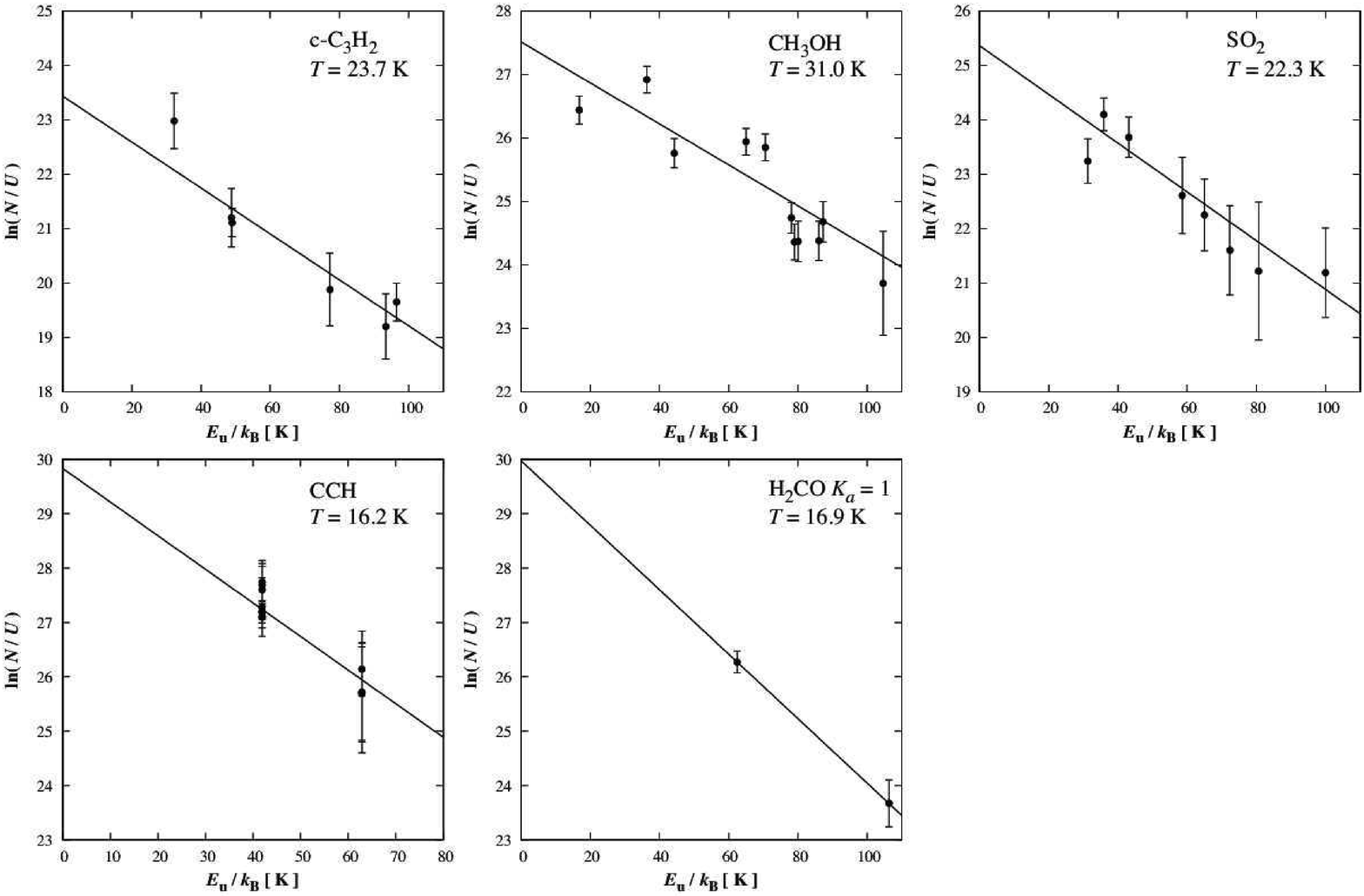}
\caption{Rotation diagram plots for c-C$_3$H$_2$, CH$_3$OH, SO$_2$, CCH, and H$_2$CO.}
\label{fig4}
\end{figure}

\clearpage
\begin{figure}
\epsscale{0.55}
\plotone{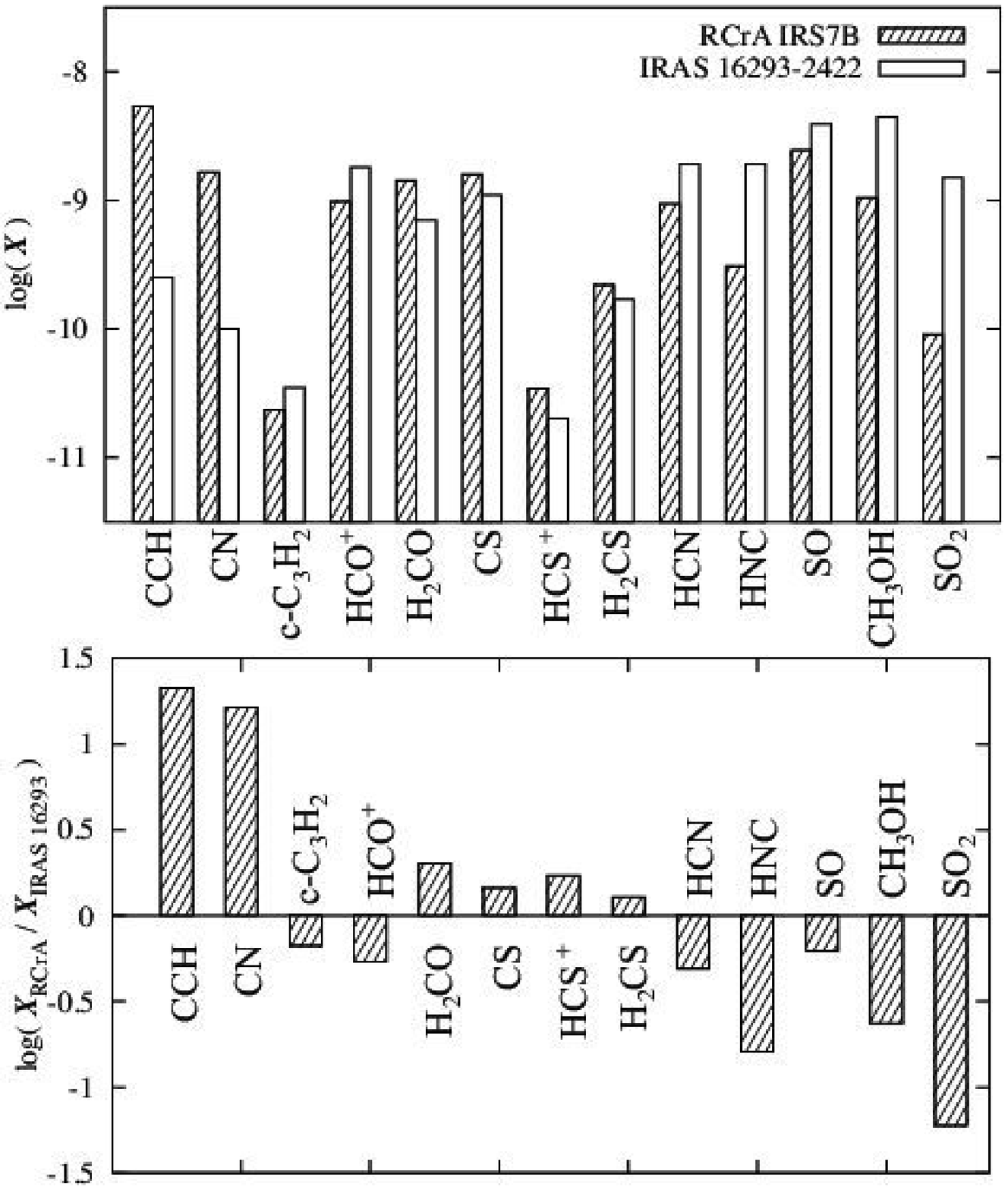}
\caption{The fractional abundances ($X$) of fundamental molecules in R~CrA~IRS7B and IRAS~16293-2422 (upper figure) and relative fractional abundances between R~CrA~IRS7B and IRAS~16293-2422 (lower figure).  The fractional abundances in IRAS~16293-2422 are taken from \citet{bla94} and \citet{dis95}.}
\label{fig5}
\end{figure}

\clearpage

%% Tables
\begin{deluxetable}{lllrrrrr}
\rotate
\tabletypesize{\small}
\tablecolumns{7}
\tablewidth{586.59421pt}
\tablecaption{Observed line parameters}
\tablehead{
\colhead{Frequency} & \colhead{Molecule}   & \colhead{Transition\qquad\qquad\qquad\qquad\qquad\qquad\qquad} & \colhead{$V_{\rm LSR}$ } & { $\Delta v$ } & \colhead{$T_{\rm mb}$ } & \colhead{ r.m.s. } & \colhead{ $\int T_{\rm mb} dv$ }\\
\colhead{GHz} & \colhead{} & \colhead{} & \colhead{km~s$^{-1}$} & \colhead{km~s$^{-1}$} & \colhead{mK} & \colhead{mK} & \colhead{K*km~s$^{-1}$} }
\startdata
332.505242 & ${\rm SO_{2}}$ & $4_{3\,1}-3_{2\,2} $ & 5.3 & 2.4 & 70.3 & 14.7 & 0.17 $\pm$ 0.06\\
334.673353 & ${\rm SO_{2}}$ & $ 8_{2\,6}-7_{1\,7} $ & 5.5 & 1.9 & 99.5 & 14.5 & 0.19 $\pm$ 0.06\\
335.096782 & ${\rm HDCO }$ & $ 5_{1\,4}-4_{1\,3} $ & 5.7 & 2.2 & 187.5 & 12.5 & 0.43 $\pm$ 0.07\\
335.582005 & ${\rm CH_{3}OH}$ & $ 7_{1}-6_{1},{\rm A^{+}}$ & 5.5 & 2.3 & 150.8 & 13.8 & 0.42 $\pm$ 0.08\\
336.948590 & ${\rm c-C_{3}H_{2}}$ & $ 4_{4\,1}-3_{1\,2} $ & 5.8 & 1.9 & 77.0 & 15.7 & 0.15 $\pm$ 0.07\\
337.060988 & ${\rm C^{17}O}$ & $ J = 3-2 $ & 5.5 & 1.8 & 3969.8 & 13.2 & 7.85 $\pm$ 0.09\\
337.396459 & ${\rm C^{34}S}$ & $ J = 7-6 $ & 5.9 & 1.7 & 333.3 & 13.2 & 0.65 $\pm$ 0.10\\
337.580147 & ${\rm ^{34}SO}$ & $ N_J = 8_{8}-7_{7} $ & 5.9 & 1.6 & 71.2 & 15.8 & 0.14 $\pm$ 0.08\\
338.083195 & ${\rm H_{2}CS}$ & $ 10_{1\,10}-9_{1\,9} $ & 5.8 & 1.9 & 83.7 & 13.8 & 0.14 $\pm$ 0.07\\
338.124502 & ${\rm CH_{3}OH}$ & $ 7_{0}-6_{0},{\rm E} $ & 5.7 & 2.2 & 252.5 & 13.8 & 0.63 $\pm$ 0.08\\
338.203988 & ${\rm c-C_{3}H_{2}}$ & $ 5_{5\,1}-4_{4\,0} $ & 5.9 & 1.5 & 82.5 & 13.8 & 0.14 $\pm$ 0.07\\
338.344628 & ${\rm CH_{3}OH}$ & $ 7_{-1}-6_{-1},{\rm E} $ & 5.7 & 2.3 & 701.5 & 13.8 & 1.87 $\pm$ 0.10\\
338.408681 & ${\rm CH_{3}OH}$ & $ 7_{0}-6_{0},{\rm A^{+}} $ & 5.6 & 2.3 & 744.2 & 13.8 & 2.10 $\pm$ 0.10\\
338.512627 & ${\rm CH_{3}OH}$ & $ 7_{4}-6_{4},{\rm A^{-}} $ & 5.5 & 3.3 & 58.5 & 16.0 & 0.12 $\pm$ 0.07\\
338.512639 & ${\rm CH_{3}OH}$ & $ 7_{4}-6_{4},{\rm A^{+}} $ & - & - & - & - & \\
338.540795 & ${\rm CH_{3}OH}$ & $ 7_{1}-6_{1},{\rm E} $ & 4.9 & 3.5 & 67.0 & 16.0 & 0.21 $\pm$ 0.08\\
338.614999 & ${\rm CH_{3}OH}$ & $ 7_{2}-6_{2},{\rm E} $ & 5.7 & 2.2 & 179.7 & 16.0 & 0.42 $\pm$ 0.10\\
338.721630 & ${\rm CH_{3}OH}$ & $ 7_{1}-6_{1},{\rm E}$ &5.4 & 2.5 & 329.8 & 16.0 & 0.85 $\pm$ 0.10\\
338.722940 & ${\rm CH_{3}OH}$ & $ 7_{-2}-6_{-2},{\rm E}$ & - & - & - & - & - \\
339.341459 & ${\rm SO }$ & $ N_J = 3_{3}-2_{3} $ & 5.6 & 2.0 & 108.3 & 15.2 & 0.25 $\pm$ 0.10\\
339.446777 & ${\rm CN }$ & $ N=3-2,J=5/2-5/2,F=3/2-3/2 $ & 5.6 & 1.7 & 59.8 & 15.2 & 0.11 $\pm$ 0.06\\
339.475904 & ${\rm CN }$ & $ N=3-2,J=5/2-5/2,F=5/2-5/2 $ & 5.7 & 1.8 & 82.3 & 15.2 & 0.16 $\pm$ 0.06\\
339.499288 & ${\rm CN }$ & $ N=3-2,J=5/2-5/2,F=7/2-5/2 $ & 5.4 & 2.3 & 47.7 & 15.2 & 0.11 $\pm$ 0.06\\
339.516635 & ${\rm CN }$ & $ N=3-2,J=5/2-5/2,F=7/2-7/2 $ & 5.8 & 1.9 & 197.7 & 15.0 & 0.40 $\pm$ 0.09\\
339.857269 & ${\rm ^{34}SO }$ & $ N_J = 8_{9}-7_{8} $ & 5.7 & 2.2 & 51.3 & 15.0 & 0.10 $\pm$ 0.08\\
339.981 & ${\rm U }$ &  & -  & - & 91.8 & 15.0 & 0.25 $\pm$ 0.08 \\
339.984 & ${\rm U }$ &  & -  & - & 115.2 & 15.0 & 0.28 $\pm$ 0.08\\
340.008126 & ${\rm CN }$ & $ N=3-2,J=5/2-3/2,F=5/2-5/2 $ & 5.7 & 2.2 & 173.8 & 18.2 & 0.41 $\pm$ 0.12\\
340.019626 & ${\rm CN }$ & $ N=3-2,J=5/2-3/2,F=3/2-3/2 $ & 5.8 & 1.7 & 169.3 & 18.2 & 0.30 $\pm$ 0.11\\
340.031549 & ${\rm CN }$ & $ N=3-2,J=5/2-3/2,F=7/2-5/2 $ & 5.7 & 2.3  & 1386.1 & 18.2 & 3.47 $\pm$ 0.10\\
340.035408 & ${\rm CN }$ & $ N=3-2,J=5/2-3/2,F=3/2-1/2 $ & 5.8 & 2.3 & 1359.2 & 18.2 & 3.32 $\pm$ 0.10\\
340.035408 & ${\rm CN }$ & $ N=3-2,J=5/2-3/2,F=5/2-3/2 $ & - & - & - & - & - \\
340.247770 & ${\rm CN }$ & $ N=3-2,J=7/2-5/2,F=7/2-5/2 $ & 5.6 & 2.7 & 2917.2 & 18.2 & 9.46 $\pm$ 0.13\\
340.247770 & ${\rm CN }$ & $ N=3-2,J=7/2-5/2,F=9/2-7/2 $ & - & - & - & - & - \\
340.248544 & ${\rm CN }$ & $ N=3-2,J=7/2-5/2,F=5/2-3/2 $ & - & - & - & - & - \\
340.261773 & ${\rm CN }$ & $ N=3-2,J=7/2-5/2F=5/2-5/2 $ & 5.6 & 2.0 & 235.0 & 18.2 & 0.42 $\pm$ 0.08\\
340.264949 & ${\rm CN }$ & $ N=3-2,J=7/2-5/2,F=7/2-7/2 $ & 5.7 & 1.9 & 256.7 & 18.2 & 0.54 $\pm$ 0.09\\
340.630692 & ${\rm HC^{18}O^{+} }$ & $ J = 4-3 $ & 5.7 & 1.6 & 193.7 & 21.0 & 0.35 $\pm$ 0.11\\
340.714155 & ${\rm SO }$ & $ N_J = 8_{7}-7_{6} $ & 5.4 & 2.7 & 831.7 & 21.0 & 2.41 $\pm$ 0.13\\
341.350229 & ${\rm HCS^{+} }$ & $ J = 8-7 $ & 5.6 & 2.5 & 59.3 & 17.8 & 0.18 $\pm$ 0.09\\
341.415639 & ${\rm CH_{3}OH }$ & $ 7_{1}-6_{1},{\rm A^{-} }$ & 5.8 & 2.5 & 135.2 & 17.8 & 0.43 $\pm$ 0.11\\
342.522128 & ${\rm D_{2}CO }$ & $ 6_{0\,6}-5_{0\,5} $ & 5.6 & 1.6 & 115.8 & 20.2 & 0.21 $\pm$ 0.12\\
342.882850 & ${\rm CS }$ & $ J=7-6 $ & 5.8 & 2.1 & 5086.7 & 20.2 & 12.38 $\pm$ 0.13\\
343.325713 & ${\rm H_{2}^{13}CO}$ & $ 5_{1\,5}-4_{1\,4} $ & 5.9 & 1.5 & 133.2 & 20.3 & 0.24 $\pm$ 0.10\\
344.200109 & ${\rm HC^{15}N }$ & $ J=4-3 $ & 5.8 & 1.5 & 130.8 & 20.2 & 0.23 $\pm$ 0.09\\
344.310612 & ${\rm SO }$ & $ N_J=8_{8}-7_{7} $ & 5.5 & 2.7 & 940.2 & 20.2 & 1.99 $\pm$ 0.13\\
345.339769 & ${\rm H^{13}CN }$ & $ J=4-3 $ & 6.0 & 2.0 & 435.2 & 15.2 & 0.95 $\pm$ 0.10\\
345.795990 & ${\rm ^{12}CO }$ & $ J=3-2 $ & - & - & 34982.8 & 16.2 & 282.26 $\pm$ 0.25\\
346.487 & ${\rm U }$ & & - & - & 66.2 & 13.3 & 0.19 $\pm$ 0.07\\
346.528481 & ${\rm SO }$ & $ N_J= 8_{9}-7_{8} $ & 5.6 & 2.5 & 940.2 & 12.7 & 2.67 $\pm$ 0.08\\
346.998344 & ${\rm H^{13}CO^{+} }$ & ${ J=4-3 }$ & 5.8 & 1.7 & 876.3 & 12.7 & 1.68 $\pm$ 0.07\\
347.740011 & ${\rm SO^{+} }$ & $ \Omega=1/2,J=15/2-13/2,f $ & 5.6 & 4.0 & 58.0 & 14.5 & 0.23 $\pm$ 0.06\\
348.115221 & ${\rm SO^{+} }$ & $ \Omega=1/2,J=15/2-13/2,e $ & 5.5 & 1.7 & 35.8 & 12.3 & 0.08 $\pm$ 0.06\\
348.211153 & ${\rm HC^{17}O^{+} }$ & $ J=4-3 $ & 6.1 & 1.0 & 64.0 & 12.3 & 0.08 $\pm$ 0.05\\
348.340814 & ${\rm HN^{13}C }$ & $ J=4-3 $ & 5.6 & 1.4 & 216.0 & 12.3 & 0.32 $\pm$ 0.06\\
348.534365 & ${\rm H_{2}CS }$ & $ 10_{1\,9}-9_{1\,8} $ & 6.0 & 1.4 & 95.8 & 13.2 & 0.13 $\pm$ 0.05\\
349.263978 & ${\rm c-C_{3}H_{2}}$ & $ 5_{5\,0}-4_{4\,1} $ & 5.8 & 1.4 & 231.5 & 12.0 & 0.37 $\pm$ 0.06\\
349.312832 & ${\rm CCH }$ & $ N = 4 - 3,J = 9/2 - 7/2,F = 4 - 4 $ & 5.7 & 1.3 & 58.8 & 12.0 & 0.08 $\pm$ 0.04\\
349.337706 & ${\rm CCH }$ & $ N = 4 - 3,J = 9/2 - 7/2,F = 5 - 4 $ & 5.3 & 2.3 & 2795.2 & 12.0 & 7.05 $\pm$ 0.09\\
349.338988 & ${\rm CCH }$ & $ N = 4 - 3,J = 9/2 - 7/2,F = 4 - 3 $ & - & - & - & - & - \\
349.399276 & ${\rm CCH }$ & $ N = 4 - 3,J = 7/2 - 5/2,F = 4 - 3 $ & 5.3 & 2.4 & 1922.0 & 12.0 & 4.89 $\pm$ 0.09\\
349.400671 & ${\rm CCH }$ & $ N = 4 - 3,J = 7/2 - 5/2,F = 3 - 2 $ & - & - & - & - & - \\
349.414643 & ${\rm CCH }$ & $ N = 4 - 3,J = 7/2 - 5/2,F = 3 - 3 $ & 5.8 & 1.9 & 97.3 & 12.0 & 0.22 $\pm$ 0.06\\
349.603614 & ${\rm CCH }$ & $ N = 4 - 3,J = 7/2 - 7/2,F = 4 - 4 $ & 5.7 & 1.5 & 124.2 & 11.8 & 0.22 $\pm$ 0.06\\
349.629770 & ${\rm CCH }$ & $ N = 4 - 3,J = 7/2 - 7/2,F = 4 - 3 $ & 5.2 & 1.6 & 73.2 & 11.8 & 0.13 $\pm$ 0.06\\
349.645137 & ${\rm CCH }$ & $ N = 4 - 3,J = 7/2 - 7/2,F = 3 - 3 $ & 5.7 & 1.8 & 60.5 & 11.8 & 0.10 $\pm$ 0.05\\
350.687730 & ${\rm CH_{3}OH }$ & $ 4_{0} - 3_{-1},{\rm E } $ & 5.6 & 2.6 & 479.0 & 18.0 & 1.54 $\pm$ 0.11\\
350.689494 & ${\rm NO }$ & $ \Omega = 1/2^{+},J = 7/2 - 5/2,F = 9/2 - 7/2 $ & - & - & - & - & - \\
350.690766 & ${\rm NO }$ & $ \Omega = 1/2^{+},J = 7/2 - 5/2,F = 7/2 - 5/2 $ & - & - & - & - & - \\
350.694772 & ${\rm NO }$ & $ \Omega = 1/2^{+},J = 7/2 - 5/2,F = 5/2 - 3/2 $ & 5.0 & 1.0 & 86.3 & 18.0 & 0.16 $\pm$ 0.09\\
350.905119 & ${\rm CH_{3}OH }$ & $ 1_{1} - 0_{0},{\rm A^{+} }$ & 5.8 & 2.1 & 499.0 & 18.0 & 1.21 $\pm$ 0.12\\
351.043524 & ${\rm NO }$ & $ \Omega = 1/2^{-},J = 7/2 - 5/2,F = 9/2 - 7/2 $ & 5.6 & 1.7 & 99.5 & 16.0 & 0.19 $\pm$ 0.09\\
351.051469 & ${\rm NO }$ & $ \Omega = 1/2^{-},J = 7/2 - 5/2,F = 7/2 - 5/2 $ & 5.2 & 2.5 & 141.0 & 16.0 & 0.32 $\pm$ 0.09\\
351.051705 & ${\rm NO }$ & $ \Omega = 1/2^{-},J = 7/2 - 5/2,F = 5/2 - 3/2 $ & - & - & - & - & - \\
351.257223 & ${\rm SO_{2} }$ & $ 5_{3\,3} - 4_{2\,2} $ & 5.4 & 2.2 & 199.5 & 16.0 & 0.45 $\pm$ 0.10\\
351.523271 & ${\rm c-C_{3}H_{2}}$ & $ 7_{3\,4} - 6_{4\,3} $ & 5.8 & 1.8 & 47.7 & 15.0 & 0.11 $\pm$ 0.07\\
351.768645 & ${\rm H_{2}CO }$ & $ 5_{1\,5} - 4_{1\,4} $ & 5.8 &2.2 & 5459.3 & 15.0 & 14.07 $\pm$ 0.11\\
351.781569 & ${\rm c-C_{3}H_{2}}$ & $ 10_{1\,10} - 9_{0\,9} $ & 5.7 & 1.9 & 137.3 & 15.0 & 0.32 $\pm$ 0.09\\
351.781569 & ${\rm c-C_{3}H_{2}}$ & $ 10_{0\,10} - 9_{1\,9} $ & - & - & - & - & - \\
351.965934 & ${\rm c-C_{3}H_{2}}$ & $ 9_{1\,8} - 8_{2\,7} $ & 5.8 & 1.9 & 66.3 & 15.0 & 0.16 $\pm$ 0.09\\
351.965939 & ${\rm c-C_{3}H_{2}}$ & $ 9_{2\,8} - 8_{1\,7} $ & - & - & - & - & - \\
352.193636 & ${\rm c-C_{3}H_{2}}$ & $ 8_{3\,5} - 7_{2\,5} $ & 6.0 & 1.2 & 45.7 & 15.3 & 0.05 $\pm$ 0.05\\
352.345 & ${\rm U }$ &  & - & - & 86.8 & 15.3 & 0.16 $\pm$ 0.08\\
353.811872 & ${\rm H_{2}^{13}CO}$ & $ 5_{0\,5} - 4_{0\,4} $ & 5.7 & 1.9 & 41.0 & 15.2 & 0.08 $\pm$ 0.06\\
354.240092 & ${\rm c-C_{3}H_{2}}$ & $ 4_{3\,2} - 3_{0\,3} $ & 5.3 & 1.8 & 60.0 & 14.5 & 0.10 $\pm$ 0.06\\
354.505477 & ${\rm HCN }$ & $ J = 4 - 3 $ & 5.7 & 3.1 & 4353.0 & 14.5 & 4.71 $\pm$ 0.11\\
355.439498 & ${\rm H^{15}NC }$ & $ J = 4 - 3 $ & 5.8 & 1.1 & 39.8 & 12.8 & 0.05 $\pm$ 0.04\\
356.734223 & ${\rm HCO^{+} }$ & $ J = 4 - 3 $ & - & - & 19635.5 & 12.8 & 56.00 $\pm$ 0.14\\
356.755190 & ${\rm SO_{2} }$ & $ 10_{4\,6} - 10_{3\,7} $ & 5.6 & 1.7 & 69.2 & 12.8 & 0.16 $\pm$ 0.06\\
357.387580 & ${\rm SO_{2} }$ & $ 11_{4\,8} - 11_{3\,9} $ & 5.7 & 1.7 & 28.2 & 11.2 & 0.05 $\pm$ 0.04\\
357.581449 & ${\rm SO_{2} }$ & $ 8_{4\,4} - 8_{3\,5} $ & 5.3 & 1.9 & 34.0 & 11.7 & 0.05 $\pm$ 0.04\\
357.671821 & ${\rm SO_{2} }$ & $ 9_{4\,6} - 9_{3\,7} $ & 5.5 & 2.3 & 78.3 & 11.7 & 0.04 $\pm$ 0.05\\
357.871456 & ${\rm D_{2}CO }$ & $ 6_{2\,4} - 5_{2\,3} $ & 6.2 & 1.0 & 57.8 & 11.7 & 0.09 $\pm$ 0.05\\
357.892442 & ${\rm SO_{2} }$ & $ 7_{4\,4} - 7_{3\,5} $ & 5.7 & 1.6 & 39.5 & 11.7 & 0.08 $\pm$ 0.05\\
357.925848 & ${\rm SO_{2} }$ & $ 6_{4\,2} - 6_{3\,3} $ & 5.4 & 2.7 & 37.0 & 11.7 & 0.09 $\pm$ 0.06\\
358.605800 & ${\rm CH_{3}OH }$ & $ 4_{1} - 3_{0},{\rm E }$ & 5.7 & 2.1 & 287.3 & 11.8 & 0.70 $\pm$ 0.08\\
360.169778 & ${\rm DCO^{+} }$ & $ J = 5 - 4 $ & 5.8 & 1.4 & 438.5 & 22.5 & 0.65 $\pm$ 0.15\\
360.618340 & ${\rm CCD }$ & $ N = 5 - 4,J = 11/2 - 9/2 $ & 5.6 & 1.3 & 179.8 & 20.0 & 0.30 $\pm$ 0.11\\ 
360.674170 & ${\rm CCD }$ & $ N = 5 - 4,J = 9/2 - 7/2 $ & 5.7 & 1.5 & 153.0 & 20.0 & 0.24 $\pm$ 0.12\\
361.852251 & ${\rm CH_{3}OH }$ & $ 8_{1} - 7_{2},{ \rm E }$ & 4.9 & 2.2 & 49.3 & 20.2 & 0.10 $\pm$ 0.08\\
362.045754 & ${\rm DCN }$ & $ J = 5 - 4 $ & 6.0 & 1.4 & 231.0 & 19.5 & 0.32 $\pm$ 0.11\\
362.630303 & ${\rm HNC }$ & $ J = 4 - 3 $ & 5.9 & 1.9 & 4519.5 & 19.0 & 9.81 $\pm$ 0.13\\
362.736048 & ${\rm H_{2}CO }$ & $ 5_{0\,5 }-4_{0\,4} $ & 5.9 & 2.0 & 3149.3 & 19.0 &6.95 $\pm$ 0.14\\
363.739820 & ${\rm CH_{3}OH }$ & $ 7_{2}-6_{1},{ \rm E }$ & 5.8 & 2.1 & 226.0 & 20.8 & 0.51 $\pm$ 0.13\\
363.945894 & ${\rm H_{2}CO }$ & $ 5_{2\,4}-4_{2\,3} $ & 5.9 & 2.1 & 510.0 & 20.8 & 1.29 $\pm$ 0.15\\
436.660979 & ${\rm CCH }$ & $ N = 5 - 4,J = 11/2 - 9/2,F = 6 - 5 $ & 5.6 & 1.7 & 2093.2 & 223.8 & 3.8 $\pm$ 1.3\\
436.661819 & ${\rm CCH }$ & $ N = 5 - 4,J = 11/2 - 9/2,F = 5 - 4 $ & - & - & - & - & - \\
436.723016 & ${\rm CCH }$ & $ N = 5 - 4,J = 9/2 - 7/2,F = 5 - 4 $ & 5.4 & 1.7 & 1435.4 & 223.8 & 2.0 $\pm$ 1.4\\
436.723910 & ${\rm CCH }$ & $ N = 5 - 4,J = 9/2 - 7/2,F = 4 - 3 $ & - & - & - & - & - \\
461.040768 & ${\rm ^{12}CO }$ & $ J = 4 - 3 $ & - & - & 35125.6 & 141.7 & 332.1 $\pm$ 1.8\\
491.968369 & ${\rm H_{2}CO }$ & $ 7_{1\,7}-6_{1\,6} $ & 5.9 & 1.7 & 1261.0 & 162.7 & 2.1 $\pm$ 0.8\\
492.160651 & ${\rm C }$ & $ ^{3}P_{1}-^{3}P_{0} $ & 5.9 & 3.9 & 9179.8 & 176.5 & 35.5 $\pm$ 1.4\\
\enddata
\label{tab1}
\end{deluxetable}

\clearpage
\begin{center}
\begin{table}[ht]
\caption{Results of the Rotation Diagram Analyses.}
\begin{tabular}{ccc}
\tableline\tableline
Molecule & $T$ (K) & $N$ (cm$^{-2}$)\\
\tableline
c-C$_3$H$_2$ & $23.7 \pm 4.7$ & $(2.4 \pm 1.7 ) \times 10^{12}$\\
CH$_3$OH & $31.0 \pm 6.8$ & $(1.1 \pm 0.6 ) \times 10^{14}$\\
SO$_2$ & $22.3 \pm 4.9$ & $(1.2 \pm 0.7 ) \times 10^{14}$\\
CCH & $16.2 \pm 3.6$ & $(2.9 \pm 1.9 ) \times 10^{14}$\\
H$_2$CO ($K_a=1$) & $16.9 ^{+5.4}_{-3.3} $ & $9.3^{+9.3}_{-4.4} \times 10^{13}$ \tablenotemark{a}\\
\tableline
\tablenotetext{a}{ A column density of H$_2$CO in the $K_{a}$ = 1 state.  }
\end{tabular}
\label{tab2}
\end{table}
\end{center}

\clearpage
\begin{deluxetable}{cccc}
\tablecolumns{4}
\tablewidth{0pt}
\tablecaption{Column Densities of Identified Molecules.}
\tablehead{
\colhead{Molecule} & \colhead{$T=15$ K} & \colhead{$T=20$ K} & \colhead{$T=25$ K}\\ 
\colhead{} & \colhead{(cm$^{-2}$)} & \colhead{(cm$^{-2}$)} & \colhead{(cm$^{-2}$)} }
\startdata
CCH & $(9.8 \pm 2.1 ) \times 10^{14} $ & $ (5.5 \pm 1.2) \times 10^{14} $ & $(4.1 \pm 0.9) \times 10^{14}$\\
CCD & $(4.5 \pm 1.6 ) \times 10^{13} $ & $ (2.1 \pm 0.7 ) \times 10^{13} $ & $(1.4 \pm 0.5 ) \times 10^{13}$\\
CN & $(2.6\pm 0.4 ) \times 10^{14} $ & $ (1.7\pm 0.3 ) \times 10^{14} $ & $(1.4 \pm 0.2) \times 10^{14}$\\
HCN \tablenotemark{a}& $(1.7 \pm 0.4 ) \times 10^{14} $ & $ (9.6 \pm 2.1) \times 10^{13} $ & $(7.1 \pm 1.5) \times 10^{13}$\\
DCN & $(1.8\pm 0.6 ) \times 10^{12} $ & $ (8.3 \pm 2.6) \times 10^{11} $ & $(5.6 \pm 1.8) \times 10^{11}$\\
HNC \tablenotemark{a}& $(5.6\pm 1.4 ) \times 10^{13} $ & $ (3.1\pm 0.8) \times 10^{13} $ & $(2.3\pm 0.6) \times 10^{13}$\\
C$^{17}$O & $(7.4\pm 1.5 ) \times 10^{15} $ & $ (4.8 \pm 1.0) \times 10^{15} $ & $(3.9\pm 0.8) \times 10^{15}$\\
NO & $(3.8\pm 0.7 ) \times 10^{14} $ & $ (2.4 \pm 0.5) \times 10^{14} $ & $(1.9\pm 0.4) \times 10^{14}$\\
HCO$^{+} \tablenotemark{a}$ & $(1.8\pm 0.4 ) \times 10^{14} $ & $ (1.0\pm 0.2) \times 10^{14} $ & $(7.4\pm 1.5) \times 10^{13}$\\
DCO$^{+}$ & $(7.1\pm 1.6 ) \times 10^{11} $ & $ (2.5\pm 0.6) \times 10^{11} $ & $(1.4\pm 0.3) \times 10^{11}$\\
H$_{2}$CO \tablenotemark{a}& $(2.8\pm 1.1 ) \times 10^{14} $ & $ (1.4\pm 0.6) \times 10^{14} $ & $(1.0\pm 0.4) \times 10^{14}$\\
HDCO & $(1.4\pm 0.4) \times 10^{13} $ & $ (7.3\pm 1.8) \times 10^{12} $ & $(5.3\pm 1.3) \times 10^{12}$\\
D$_{2}$CO & $(7.8\pm 1.8 ) \times 10^{12} $ & $ (4.3\pm 0.8) \times 10^{12} $ & $(2.2 \pm 0.5 ) \times 10^{12}$\\
CS & $(4.3\pm 0.9 ) \times 10^{14} $ & $ (1.6\pm 0.3 ) \times 10^{14} $ & $(9.6\pm 1.9 ) \times 10^{13}$\\
HCS$^{+}$ & $(1.1\pm 0.4 ) \times 10^{13} $ & $ (3.5\pm 1.4 ) \times 10^{12} $ & $(1.9\pm 0.8) \times 10^{12}$\\
H$_{2}$CS & $(9.5\pm 4.2 ) \times 10^{13} $ & $ (2.2\pm 0.7) \times 10^{13} $ & $(1.0\pm 0.3) \times 10^{13}$\\
SO & $(8.4\pm 1.0 ) \times 10^{14} $ & $ (2.5\pm 0.3) \times 10^{14} $ & $(1.3\pm 0.2) \times 10^{14}$\\
SO$^{+}$ & $(1.1\pm 0.2 ) \times 10^{13} $ & $ (3.9\pm 0.8) \times 10^{12} $ & $(2.2\pm 0.5) \times 10^{12}$\\
C & $(9.1\pm 1.9 ) \times 10^{17} $ & $ (7.8\pm 1.6) \times 10^{17} $ & $(7.4\pm 1.5) \times 10^{17}$\\
(CH$_3$)$_2$O \tablenotemark{b} & $< 6.2 \times 10^{12} $ & $ < 4.8 \times 10^{12} $ & $< 4.4 \times 10^{12}$\\
HOC$^{+}$ \tablenotemark{b} & $< 2.6 \times 10^{11} $ & $ < 1.4 \times 10^{11} $ & $< 1.0 \times 10^{11}$\\
CO$^{+}$ \tablenotemark{b} & $< 3.6 \times 10^{11} $ & $ < 2.3 \times 10^{11} $ & $< 1.8 \times 10^{11}$\\
\enddata
\tablenotetext{a}{The spectral lines of the $^{13}$C species is used for evaluation of the column density, where the $^{12}$C/$^{13}$C ratio is assumed to be 60.  }
\tablenotetext{b}{The upper limit to the column density is estimated from the 3$\sigma$ upper limit of the integrated intensity assuming the line width to be 2.0~km/s.}
\label{tab3}
\end{deluxetable}

\clearpage
\begin{deluxetable}{cccc}
\tablecolumns{4}
\tablewidth{0pt}
\tablecaption{Fractional Abundances Relative to H$_2$.}
\tablehead{
\colhead{Molecule} & \colhead{$T=15$ K} & \colhead{$T=20$ K} & \colhead{$T=25$ K} }
\startdata
CCH & $(6.3\pm 1.8 ) \times 10^{-9} $ & $ (5.3\pm 1.5) \times 10^{-9} $ & $(4.8\pm 1.4) \times 10^{-9}$\\
CCD & $(2.8\pm 1.1 ) \times 10^{-10} $ & $ (2.0\pm 0.8) \times 10^{-10} $ & $(1.7\pm 0.7) \times 10^{-10}$\\
CN & $(1.7\pm 0.4 ) \times 10^{-9} $ & $ (1.6\pm 0.4) \times 10^{-9} $ & $(1.6\pm 0.4) \times 10^{-9}$\\
HCN & $(1.1\pm 0.3 ) \times 10^{-9} $ & $ (9.3\pm 2.7) \times 10^{-10} $ & $(8.5\pm 2.5) \times 10^{-10}$\\
DCN & $(1.1\pm 0.4 ) \times 10^{-11} $ & $ (8.1\pm 3.0) \times 10^{-12} $ & $(6.6\pm 2.5) \times 10^{-12}$\\
HNC & $(3.6\pm 1.2 ) \times 10^{-10} $ & $ (3.1\pm 1.0 ) \times 10^{-10} $ & $(2.8 \pm 0.9 ) \times 10^{-10}$\\
NO & $(2.4 \pm 0.7 ) \times 10^{-9} $ & $ (2.3 \pm 0.6 ) \times 10^{-9} $ & $(2.2 \pm 0.6 ) \times 10^{-9}$\\
HCO$^{+}$ & $(1.1\pm 0.3 ) \times 10^{-9} $ & $ (9.7\pm 2.7) \times 10^{-10} $ & $(8.8 \pm 2.5) \times 10^{-10}$\\
DCO$^{+}$ & $(4.5 \pm 1.4 ) \times 10^{-12} $ & $ (2.5 \pm 0.7 ) \times 10^{-12} $ & $(1.6 \pm 0.5) \times 10^{-12}$\\
H$_{2}$CO & $(1.8 \pm 0.8 ) \times 10^{-9} $ & $ (1.4 \pm 0.6) \times 10^{-9} $ & $(1.2\pm 0.5) \times 10^{-9}$\\
HDCO & $(9.0\pm 2.9 ) \times 10^{-11} $ & $ (7.1 \pm 2.3) \times 10^{-11} $ & $(6.3 \pm 2.0 ) \times 10^{-11}$\\
D$_{2}$CO & $(5.0\pm 1.5) \times 10^{-11} $ & $ (4.2\pm 1.2) \times 10^{-11} $ & $(2.6\pm 0.8) \times 10^{-11}$\\
CH$_3$OH & $(6.8 \pm 3.8 ) \times 10^{-10} $ & $ (1.0 \pm 0.6 ) \times 10^{-9} $ & $(1.3 \pm 0.7) \times 10^{-9}$\\
c-C$_3$H$_2$ & $(1.5 \pm 1.1 ) \times 10^{-11} $ & $ (2.3 \pm 1.7 ) \times 10^{-11} $ & $(2.9 \pm 2.0 ) \times 10^{-11}$\\
CS & $(2.8 \pm 0.8 ) \times 10^{-9} $ & $ (1.6 \pm 0.5 ) \times 10^{-9} $ & $(1.1 \pm 0.3) \times 10^{-9}$\\
HCS$^{+}$ & $(6.8\pm 3.0 ) \times 10^{-11} $ & $ (3.4\pm 1.5) \times 10^{-11} $ & $(2.3\pm 1.0) \times 10^{-11}$\\
H$_{2}$CS & $(6.0\pm 3.0 ) \times 10^{-10} $ & $ (2.2 \pm 0.8) \times 10^{-10} $ & $(1.2\pm 0.5) \times 10^{-10}$\\
SO & $(5.4\pm 1.3 ) \times 10^{-9} $ & $ (2.4 \pm 0.6) \times 10^{-9} $ & $(1.5 \pm 0.4) \times 10^{-9}$\\
SO$^{+} $ & $(7.2\pm 2.1 ) \times 10^{-11} $ & $ (3.8 \pm 1.1) \times 10^{-11} $ & $(2.6 \pm 0.8) \times 10^{-11}$\\
SO$_2$& $(7.9\pm 4.9 ) \times 10^{-11} $ & $ (1.2\pm 0.8) \times 10^{-10} $ & $(1.5\pm 0.9) \times 10^{-10}$\\
C & $(5.8\pm 1.7 ) \times 10^{-6} $ & $ (7.5\pm 2.2) \times 10^{-6} $ & $(8.8\pm 2.5) \times 10^{-6}$\\
(CH$_3$)$_2$O & $< 3.9 \times 10^{-11} $ & $ < 4.7 \times 10^{-11} $ & $< 5.2 \times 10^{-11}$\\
HOC$^{+}$ & $< 1.6 \times 10^{-12} $ & $ < 1.4 \times 10^{-12} $ & $< 1.2 \times 10^{-12}$\\
CO$^{+}$ & $< 2.3 \times 10^{-12} $ & $ < 2.2 \times 10^{-12} $ & $< 2.2 \times 10^{-12}$\\
\enddata
\label{tab4}
\end{deluxetable}

\clearpage
\begin{center}
\begin{table}[ht]
\caption{Deuterium Fractionation Ratios.}
\begin{tabular}{cccc}
\tableline\tableline
Molecule & $T=15$ K & $T=20$ K & $T=25$ K \\
\tableline
CCD / CCH & $0.045 \pm 0.019 $ & $ 0.038 \pm 0.016 $ & $ 0.035 \pm 0.014 $\\
HDCO / H$_2$CO & $ 0.050 \pm 0.024 $ & $ 0.050 \pm 0.024 $ & $ 0.052 \pm 0.025 $\\
D$_2$CO / H$_2$CO & $ 0.028 \pm 0.013 $ & $ 0.030 \pm 0.013 $ & $ 0.0212 \pm 0.010 $\\
DCN / HCN & $ 0.010 \pm 0.003 $ & $ 0.009 \pm 0.003 $ & $  0.0008 \pm 0.002 $\\
DCO$^{+}$ / HCO$^{+}$ & $ 0.0040 \pm 0.0012 $ & $ 0.0025 \pm 0.0006 $ & $ 0.0018 \pm 0.0004 $\\
\tableline
\end{tabular}
\label{tab5}
\end{table}
\end{center}

\end{document}